\documentclass[12pt]{article}
\pdfoutput=1
\usepackage{subfigure}
\usepackage{amssymb,amsmath}
\usepackage{graphicx}
\usepackage{setspace,caption}
\usepackage{lipsum}
\usepackage{color}
\usepackage[colorlinks=true
,urlcolor=blue
,citecolor=blue
,linkcolor=blue
,pagecolor=blue
,linktocpage=true
,pdfproducer=medialab
]{hyperref}
\usepackage[a4paper]{geometry}
\usepackage{placeins}
\usepackage{cite}
\usepackage{cancel}

\makeatletter \renewcommand{\@dotsep}{10000} \makeatother

\def\tst{\tilde t}
\def\ttau{\tilde \tau}

\mathchardef\mhyphen="2D

\newcommand{\beq}{\begin{equation}}
\newcommand{\eeq}{\end{equation}}
\newcommand{\bea}{\begin{eqnarray}}
\newcommand{\eea}{\end{eqnarray}}

\begin{document}

\begin{titlepage}
\pagestyle{empty}

%\begin{flushright}
%FTPI-MINN-16/24
%\end{flushright}

\vspace*{0.2in}
\begin{center}
{\Large \bf    $b-\tau$ Yukawa Unification in SUSY SU(5)  with Mirage Mediation: LHC and Dark Matter Implications
  }\\
\vspace{1cm}
{\bf  Shabbar Raza$^{a,}$\footnote{E-mail: shabbar.raza@fuuast.edu.pk}, Qaisar Shafi$^{b,}$\footnote{E-mail: shafi@bartol.udel.edu} and
Cem Salih $\ddot{\rm U}$n$^{c,}\hspace{0.05cm}$\footnote{E-mail: cemsalihun@uludag.edu.tr}}
\vspace{0.5cm}

{\it
$^a$Department of Physics, Federal Urdu University of Arts, Science and Technology, Karachi 75300, Pakistan \\
$^b$Bartol Research Institute, Department of Physics and Astronomy,\\
University of Delaware, Newark, DE 19716, USA \\
$^c$Department of Physics, Bursa Uluda\~{g} University, 16059 Bursa, Turkey
}
\end{center}

%\vspace{0.5cm}
\begin{abstract}
\noindent
We consider a class of $b-\tau$ Yukawa unified Supersymmetric (SUSY) $SU(5)$ GUTs, in which the asymptotic gaugino $M_{1,2,3}$ masses are generated through a combination of gravity and mirage mediated supersymmetry breaking. Due to the contributions from mirage mediation, $M_{3}$ is always lighter than $M_{1}$ and $M_{2}$, and consequently for the range of asymptotic masses considered, the gluino mass $m_{\tilde{g}}$ at low scale is bounded from above at about 4 TeV. We realize two different regions, one in which the MSSM $\mu-$term is less than about 3 TeV. This region yields a stop mass up to 5 TeV, and the stop mass is nearly degenerate with the LSP neutralino for mass around 0.8 to 1.7 TeV. A stau mass can be realized up to about 5 TeV, and the stau mass is approximately degenerate with the LSP neutralino for mass around 2 to 3 TeV. In addition, an A-funnel solution with $m_{A}\approx 2m_{\tilde{\chi}_{1}^{0}}$ and $m_{\tilde{\chi}_{1}^{0}}\sim 700-900$ GeV is realized. These three cases yield LSP dark matter abundance in accordance with observations.
A second region, on the other hand, arises for $m_{\tilde{g}}\lesssim 1.1m_{\tilde{\chi}_{1}^{0}}$. The $\mu-$term is rather large ($\gtrsim 20$ TeV), and the LSP neutralino is a bino-wino mixture. The gluino mass ($\sim 0.8-1.2$ TeV) is nearly degenerate with the LSP neutralino mass and hence, the gluino-neutralino coannihilation processes play a role in reducing the relic abundance of LSP neutralino down to ranges allowed by the current WMAP measurements. The two regions above can be distinguished through the direct detection experiments. The first region with relatively low $\mu$ values yields Higgsino-like DM, whose scattering on the nucleus typically has a large cross-section. We find that such solutions are still allowed by the current results from the LUX experiment, and they will be severely tested by the LUX-Zeplin (LZ) experiment. The second region contains bino-wino DM whose scattering cross-section is relatively low. These solutions are harder to rule out in the foreseeable future.
\end{abstract}

\end{titlepage}

%\baselineskip 36pt

%%%%%%%%%%%%%%%%%%%%%%%%%%s
% Main body
%%%%%%%%%%%%%%%%%%%%%%%%%%

\section{Introduction}
\label{sec:intro}
Third family Yukawa Unification (YU) at $M_{GUT}$, such as $t-b-\tau$ YU in SUSY SO(10) \cite{big-422,bigger-422} and
$b-\tau$ YU in SUSY SU(5) \cite{Chattopadhyay:2001mj,Baer:2012by,Gogoladze:2010fu} have attracted a fair amount of attention in recent years. A characteristic feature of these models is the presence of threshold corrections generated via finite one loop diagrams \cite{Hall:1993gn}, which play an essential role in realizing YU or quasi-YU at $M_{GUT}$. These threshold corrections can provide a deeper probe of the underlying SUSY spectrum compared to the more straightforward requirement of gauge coupling unification which can be realized with a SUSY scale on the order of 1- 10 TeV. For instance, imposing the constraint that the relic  LSP neutralino saturates the dark matter (DM) abundances can yield stringent constraints on the YU spectrum. As an example $t-b-\tau$ YU with universal gaugino masses at $M_{GUT}$ fail to satisfy the DM constraint \cite{Baer:2008jn}. This can be remedied by allowing non-universality in the gaugino sector which leads to a variety of co-annihilation scenarios with the desired DM abundance \cite{Raza:2014upa,Dar:2011sj,Chattopadhyay:2001va}.

Despite no experimental evidence thus far for its existence \cite{Aad:2012tfa,CMS:2017rli}, SUSY remains a compelling extension of the highly successful Standard Model (SM.) SUSY GUTs, in particular, possess a number of attractive features. These include an attractive resolution of the gauge hierarchy problem, compelling DM candidates, and the ability to resolve some other experimental challenges faced by the SM including the muon anomalous magnetic moment. Inspired by these and related considerations, we propose in this paper to explore the low energy implications of a  SUSY SU(5) model with $b-\tau$ YU. A new feature here is our assumption that the soft SUSY breaking terms in the SU(5) sector arise from a combination of  gravity and mirage mediated SUSY breaking terms, which yields non-universal
gaugino masses and trilinear scalar interaction terms in the soft SUSY breaking (SSB) terms in the Lagrangian.

Following the analyses presented in \cite{Baer:2012by}, we explore the SUSY mass spectrum, taking into account the current LHC analyses, and the current and near future constraints from the DM experiments. Note that the non-universality in the gaugino masses generated by the mirage mediation mechanism \cite{Choi:2005uz} differentiates this study from the previous studies \cite{Baer:2012by}. 

The remainder of the paper is organized as follows: Section \ref{sec:model} briefly summarizes the model and the relevant free parameters. 
We describe in Section~\ref{sec:scan} the scanning procedure in generating data along with the ranges of the fundamental parameters, and the current experimental constraints employed in our analyses. The impact of  $b-\tau$ YU condition on low energy phenomenology is described in Section \ref{sec:btauYU}. After discussing the SUSY mass spectrum in Section \ref{sec:mass}, we continue in Section \ref{sec:DM} with the DM implications in light of the current DM experiments. Finally, we summarize our findings in Section \ref{sec:conc}.

\section{Model Description}
\label{sec:model}

The SU(5) model with b-$\tau$ YU has been extensively studied in the past. An important difference
here from previous studies is the incorporation of SUSY breaking contributions from mirage
mediation to the three MSSM gaugino masses. This modifies the gaugino masses as follows \cite{Choi:2005uz}: 
\begin{equation}
M_{i}=\left(1+\dfrac{g_{G}^{2}b_{i}\alpha}{16\pi^{2}}\log\left(\dfrac{M_{{\rm Pl}}}{m_{3/2}} \right) \right)M_{1/2},
\label{eq:gaugemasses}
\end{equation}
where $g_{G}$ denotes the unified gauge coupling at $M_{GUT}$ and $b_{i}$ denote the $\beta -$functions for $U(1)_{Y}$, $SU(2)_{L}$ and $SU(3)_{c}$ with $b_{1}=33/5$, $b_{2}=1$, $b_{3}=-1$, and $\alpha$ is a parameter representing the ratio of anomaly mediation  to moduli mediation. Note that with $b_{3}$ negative, $M_{3}$ is lighter than the other gauginos ($M_{3} < M_{1,2} $) at the GUT scale, which can lead to lighter gluinos at the low scale. 
In addition, the fundamental parameter space contains the following:

\begin{equation}
m_{10},m_{5},m_{H_{d}},m_{H_{u}}, A_{t}, A_{b}=A_{\tau}, \tan\beta,
\end{equation}
where $m_{10}$ and $m_{5}$ represent common mass for $10$ and $\bar 5$ representations respectively, and $m_{H_{d}}$, and $m_{H_{u}}$ belong to $\bar 5_{H}$ and $5_{H}$ of $SU(5)$. $A_{t}$ and $A_{b}=A_{\tau}$ are the trilinear scalar couplings for top quark, bottom quark and tau lepton, and $\tan\beta \equiv v_{u}/v_{d}$, where $v_{u,d}$ are vacuum expectation values (VEVs) of the MSSM Higgs doublets, $H_{u}$ and $H_{d}$, respectively.

\noindent Previous numerical studies have shown that the threshold corrections from the SUSY particles are crucial in realizing YU consistent with the observed quark and lepton masses for the third family. The finite correction to the b-quark Yukawa coupling in particular, is required to be appropriately large and negative \cite{Gogoladze:2010fu}. The overall threshold correction to $y_{b}$ is approximately given by \cite{Pierce:1996zz}

\begin{equation}
\delta_{y_{b}}^{{\rm finite}} \approx \dfrac{g_{3}^{2}}{12\pi^{2}}\dfrac{\mu m_{\tilde{g}}\tan\beta}{m_{\tilde{b}}^{2}}+\dfrac{y_{t}^{2}}{32\pi^{2}}\dfrac{\mu A_{t}\tan\beta}{m_{\tilde{t}^{2}}}~~.
\label{eq:deltab}
\end{equation}

In Ref. \cite{Baer:2012by} it is shown that models consistent with $10\%$ or better $b-\tau$ YU can be divided into two groups on the basis of $\tan\beta$. Models with large $\tan\beta \sim 30-60$ can accomodate $m_{h}$ around 125 GeV, while models with low $\tan\beta \sim 2-10$ find it problematic to do so. This  can be understood, if one considers the loop corrections to the SM-like Higgs boson mass, which is \cite{Carena:2012mw}

\begin{equation*}
\Delta m_{h}^{2}\simeq \dfrac{m_{t}^{4}}{16\pi^{2}v^{2}\sin^{2}\beta}\dfrac{\mu A_{t}}{M^{2}_{{\rm SUSY}}}\left[\dfrac{A_{t}^{2}}{M^{2}_{{\rm SUSY}}}-6 \right]+
\end{equation*}
\begin{equation}\hspace{1.4cm}
\dfrac{y_{b}^{4}v^{2}}{16\pi^{2}}\sin^{2}\beta\dfrac{\mu^{3}A_{b}}{M^{4}_{{\rm SUSY}}}+\dfrac{y_{\tau}^{4}v^{2}}{48\pi^{2}}\sin^{2}\beta \dfrac{\mu^{3}A_{\tau}}{m_{\tilde{\tau}}^{4}}~~.
\label{eq:higgscor}
\end{equation}

The terms in the second line of Eq.(\ref{eq:higgscor}) can be neglected for low $\tan\beta$ values, while they can provide minor contributions if $\tan\beta$ is moderate or large. For small $\tan\beta$, the main contribution to the Higgs boson mass comes from the stop sector, and the results presented in \cite{Baer:2012by} show that the stop contribution is not sufficient. One needs, in addition, the contributions from the sbottom and stau sectors as well, which can be achieved with moderate or large $\tan\beta$. We will discuss such results and the impact of the non-universal gauginos on the low energy sparticle spectrum and compare them with the previous analyses in Section \ref{sec:mass}. 

\section{Scanning Procedure and Experimental Constraints}
\label{sec:scan}
In this section  we summarize our scanning procedure and the experimental constraints employed in our analyses. We {use} the ISAJET~7.85 package~\cite{ISAJET} to perform the random scans. 
In ISAJET, the weak scale values of the gauge and third
 generation Yukawa couplings are evolved to
 $M_{\rm GUT}$ via the MSSM RGEs
 in the $\overline{DR}$ regularization scheme.
We do not strictly enforce the unification condition
 $g_3=g_1=g_2$ at $M_{\rm GUT}$, since a few percent deviation
 from unification can be assigned to the unknown GUT-scale threshold
 corrections~\cite{Hisano:1992jj}. {In our scan, on the other hand, $g_{3}$ is not allowed to deviate from the unification by more than about $3\%$.}
With the boundary conditions given at $M_{\rm GUT}$,
 all the SSB parameters, along with the gauge and Yukawa couplings,
 are evolved back to the weak scale $M_{\rm Z}$.

In evaluating Yukawa couplings, the SUSY threshold
 corrections~\cite{Pierce:1996zz} are taken into account
 at the common scale $M_{\rm SUSY}= \sqrt{m_{\tst_L}m_{\tst_R}}$.
The entire parameter set is iteratively run between
 $M_{\rm Z}$ and $M_{\rm GUT}$ using the full two-loop RGEs
 until a stable solution is obtained.
To better account for the leading-log corrections, one-loop step-beta
 functions are adopted for the gauge and the Yukawa couplings, and
 the SSB parameters $m_i$ are extracted from RGEs at appropriate scales {as}
 $m_i=m_i(m_i)$.
The RGE-improved one-loop effective potential is minimized
 at an optimized scale $M_{\rm SUSY}$, which effectively
 accounts for the leading two-loop corrections.
The full one-loop radiative corrections are incorporated
 for all sparticles.
 
\noindent Note that we set $\mu > 0$ and  use $m_t = 173.3\, {\rm GeV}$  \cite{:2009ec}.
Our results are not too sensitive to one
 or two sigma variations in the value of $m_t$  \cite{bartol2}.
We also use $m_b^{\overline{DR}}(M_{\rm Z})=2.83$ GeV
 which is hard-coded into ISAJET.  
%%%%%%%

%%%%%%%%%%%%%%%%%%%%

The fundamental parameters defined earlier in Section~\ref{sec:model}, are restricted as follows:
\begin{eqnarray}
\label{parameterRange}
0 \leq  m_{10},m_{5},m_{H_{d}},m_{H_{u}}  \leq 20~ {\rm TeV} \nonumber \\
0 \leq  M_{1/2}  \leq 5~ {\rm TeV} \nonumber \\
3.5 \leq  \alpha  \leq 6~ \nonumber \\
1 \leq m_{3/2} \leq 100~{\rm TeV} \\
-3 \leq A_{t}/m_{10} \leq 3  \nonumber \\
-20 \leq A_{b\tau}/m_{5} \leq 20 \nonumber \\
1.2 \leq \tan\beta \leq 60 .\nonumber 
\end{eqnarray}

It should be noted that the requirement of radiative electroweak symmetry breaking (REWSB)~\cite{Ibanez:1982fr} puts an important theoretical
 constraint on the parameter space.
Another important constraint comes from limits on the cosmological
 abundance of stable charged {particles}~\cite{Beringer:1900zz}.
This excludes regions in the parameter space where charged
 SUSY particles, such as $\ttau_1$ or $\tst_1$,
 become the LSP.

In scanning the parameter space, we employ the Metropolis-Hastings algorithm as described in \cite{Belanger:2009ti}. The data points collected all satisfy the requirement of REWSB, with the neutralino being the LSP. In addition, after collecting the data, we impose the mass bounds on all the sparticles \cite{Agashe:2014kda}, and the constraints from rare decay processes $B_{s}\rightarrow \mu^{+}\mu^{-} $ \cite{Aaij:2012nna}, $b\rightarrow s \gamma$ \cite{Amhis:2012bh}, and $B_{u}\rightarrow \tau\nu_{\tau}$ \cite{Asner:2010qj}. We also require the relic abundance of the LSP neutralino to satisfy the WMAP bound within $5\sigma$ \cite{Hinshaw:2012aka}. More explicitly, we set

\begin{eqnarray}
m_h  = 123-127~{\rm GeV}~~&
\\
m_{\tilde{g}}\geq 2.1~{\rm TeV} (600~{\rm GeV~if~} m_{\tilde{g}}\lesssim 1.1m_{\tilde{\chi}_{1}^{0}})\\
0.8\times 10^{-9} \leq{\rm BR}(B_s \rightarrow \mu^+ \mu^-)
  \leq 6.2 \times10^{-9} \;(2\sigma)~~&&
\\
2.99 \times 10^{-4} \leq
  {\rm BR}(b \rightarrow s \gamma)
  \leq 3.87 \times 10^{-4} \; (2\sigma)~~&&
\\
0.15 \leq \frac{
 {\rm BR}(B_u\rightarrow\tau \nu_{\tau})_{\rm MSSM}}
 {{\rm BR}(B_u\rightarrow \tau \nu_{\tau})_{\rm SM}}
        \leq 2.41 \; (3\sigma)~~&&
\\
 0.0913 \leq \Omega_{\rm CDM}h^2 (\rm WMAP9) \leq 0.1363   \; (5\sigma)~~&&.
%\\
% 2.1 \times 10^{-10} \leq \Delta a_{\mu}
%  \leq 50.1 \times 10^{-10} \; (3\sigma)~~&\cite{BNL}&
%\labels{constraints}
\end{eqnarray}

{Note that we employ the WMAP bound on the relic abundance of the neutralino LSP within $5\sigma$, which used to coincide with the results from the Planck satellite within $5\sigma$ \cite{Ade:2015xua}. However, new results have recently been released by the Planck satellite, as we were completing out our analyses. These results provide a more restictive bound on the relic abundance of the neutralino LSP, namely $0.114 \leq \Omega h^{2}\leq 0.126~(5\sigma)$ \cite{Akrami:2018vks}. Considering the large uncertainties in theoretical calculations of the relic abundance of the neutralino LSP, we continue to employ the less restrictive WMAP bound in our analyses. On the other hand, we will present some benchmark points to exemplify and summarize our results before concluding our discussion. These benchmark points are compatible with the latest results from the Planck satellite.}

In addition to the constraints mentioned above, we quantify $b-\tau$ YU with the parameter $R_{b\tau}$ defined as \cite{Raza:2014upa}

\begin{equation}
R_{b\tau}\equiv \dfrac{{\rm Max}(y_{b},y_{\tau})}{{\rm Min}(y_{b},y_{\tau})},
\end{equation}

\noindent where $R_{b\tau}=1$ means perfect $b-\tau$ YU. However, considering the various uncertainties we label  solutions as compatible with $b-\tau$ YU for $R_{b\tau}\leq 1.1$.

\section{Parameter Space for $\mathbf{b-\tau}$ YU}
\label{sec:btauYU}

\begin{figure}[h!]
\centering
\subfigure{\includegraphics[scale=1.1]{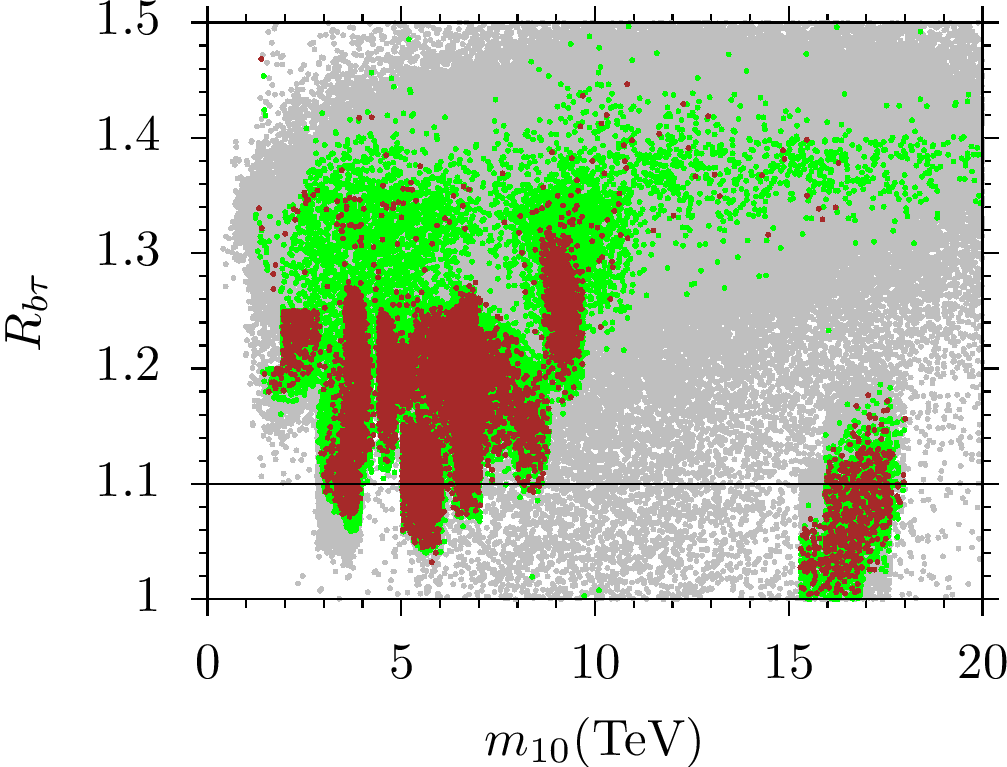}}%
\subfigure{\includegraphics[scale=1.1]{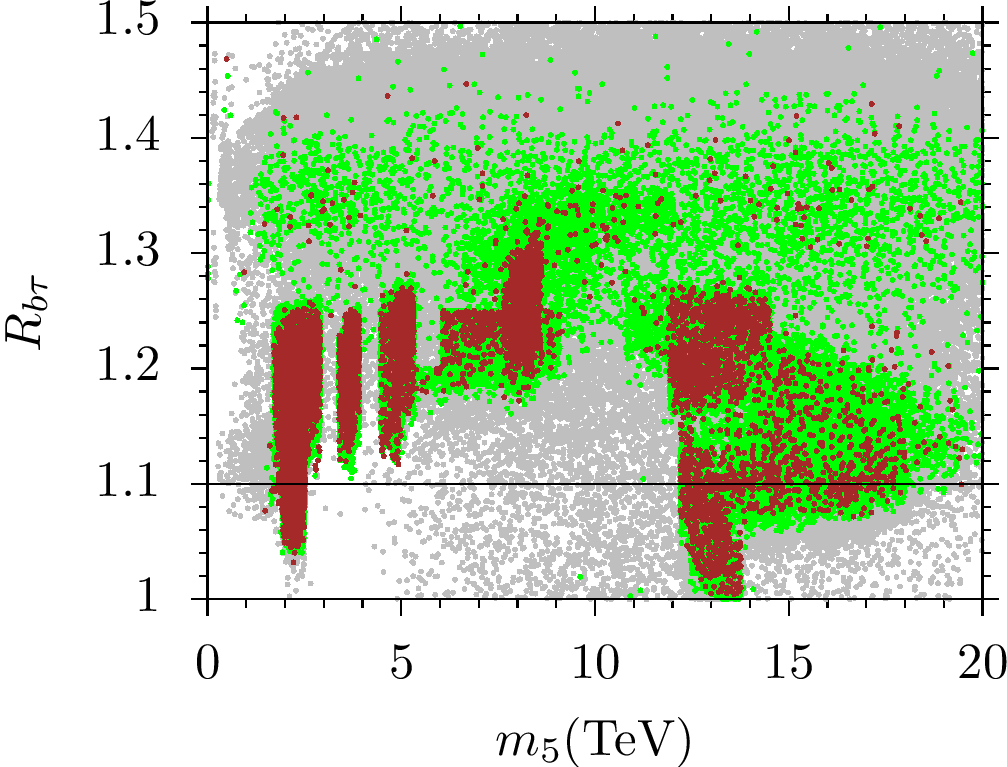}}
\subfigure{\includegraphics[scale=1.1]{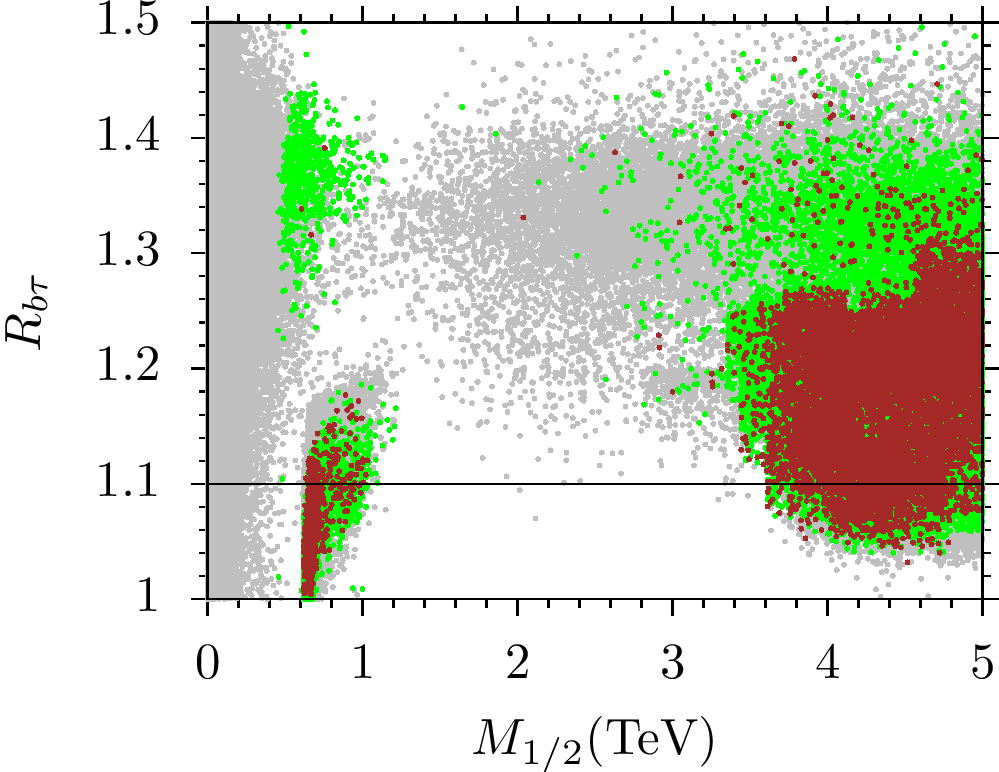}}%
\subfigure{\includegraphics[scale=1.1]{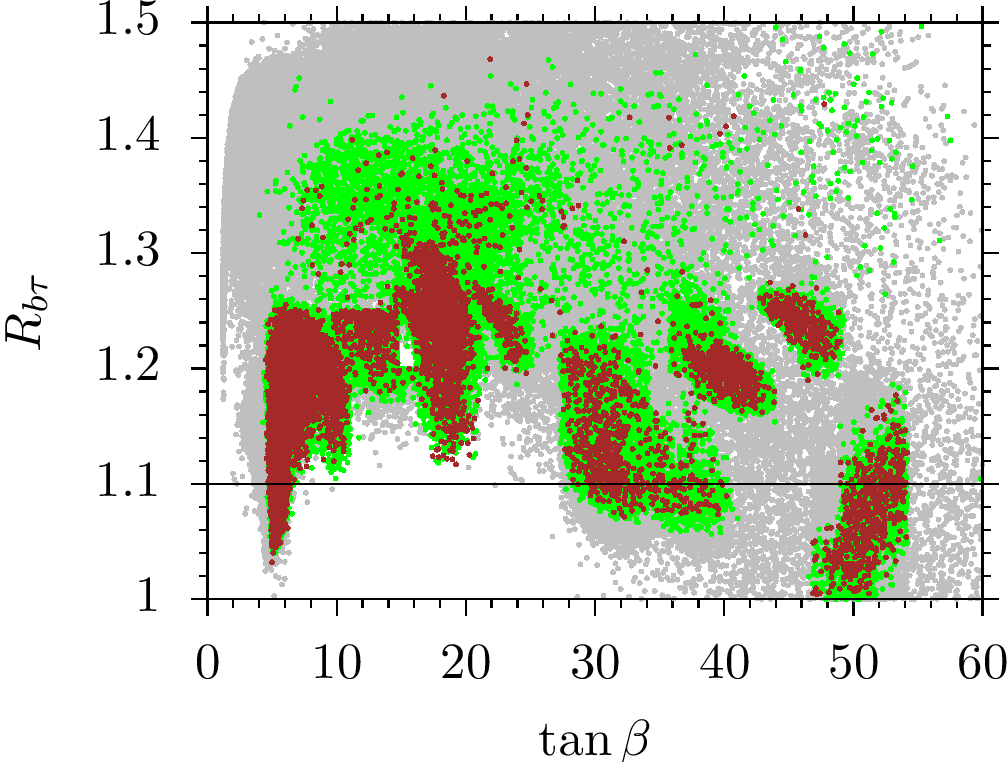}}
\caption{Plots in the $R_{b\tau}-m_{10}$, $R_{b\tau}-m_{5}$, $R_{b\tau}-M_{1/2}$, and $R_{b\tau}-\tan\beta$ planes. All points are consistent with the REWSB and LSP neutralino LSP conditions. Green points satisfy the mass bounds and the constraints from rare $B-$meson decays. Brown points are a subset of green, and they are compatible within $5\sigma$ with the WMAP bound on the relic density of the LSP neutralino. The horizontal line indicates the regions yielding $R_{b\tau}=1.1$, below which lie the solutions most consistent with the $b-\tau$ YU condition.}
\label{fig1}
\end{figure}

\begin{figure}[h!]
\centering
\subfigure{\includegraphics[scale=1.1]{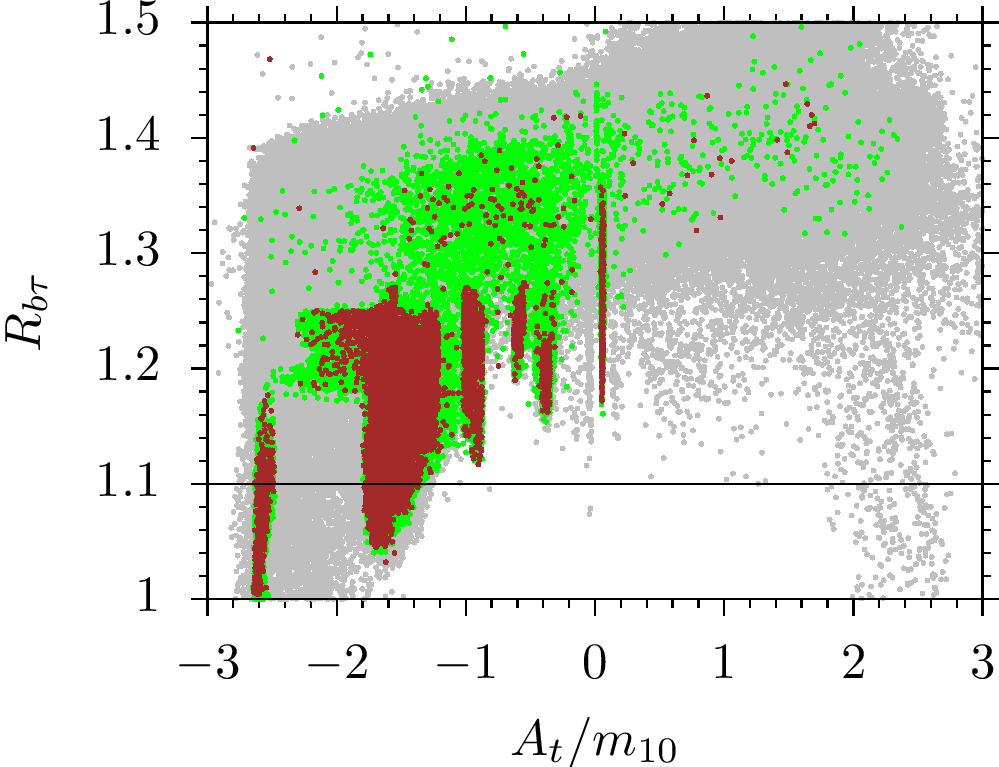}}
\subfigure{\includegraphics[scale=1.1]{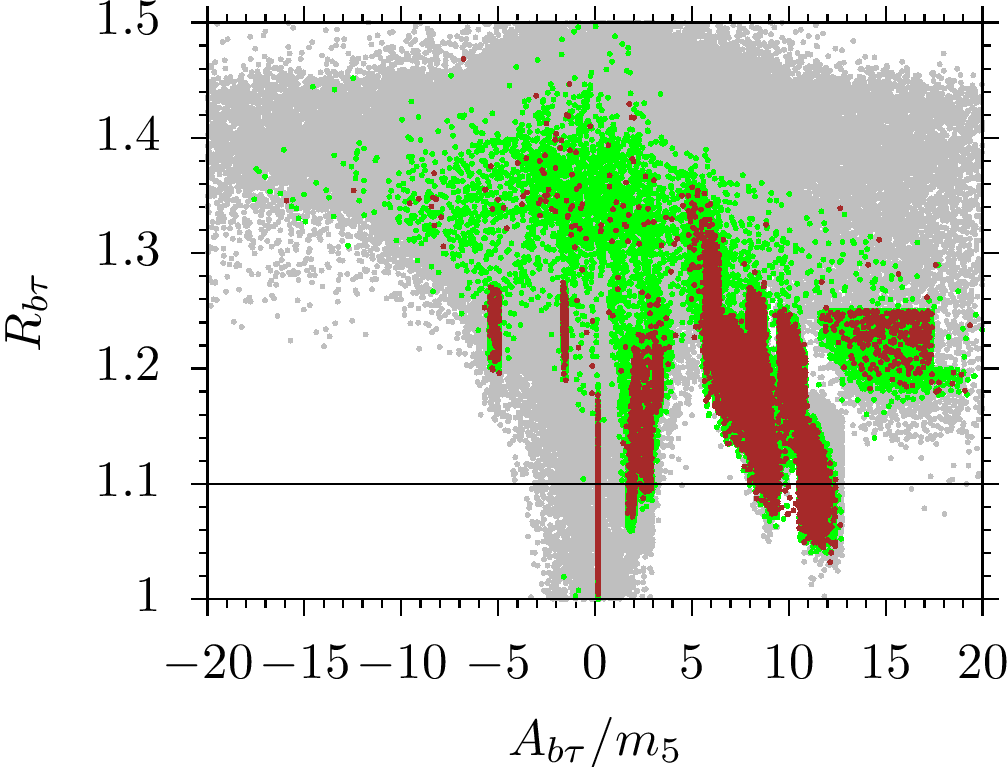}}
\subfigure{\includegraphics[scale=1.1]{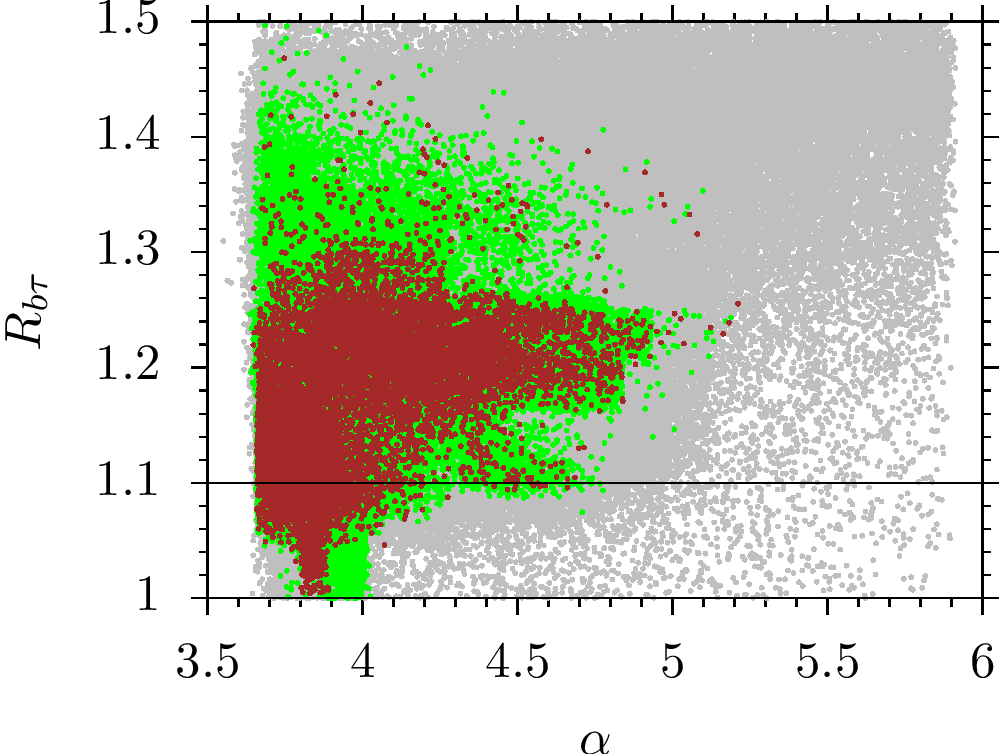}}
\subfigure{\includegraphics[scale=1.1]{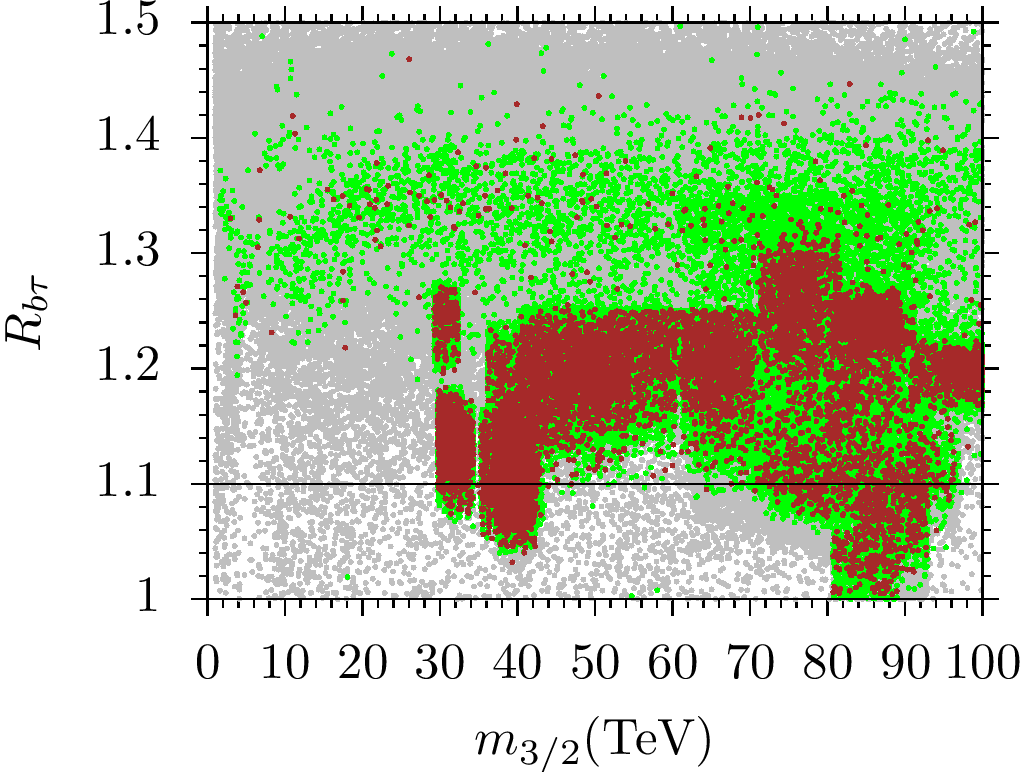}}
\caption{Plots in the $R_{b\tau}-A_{t}/m_{10}$, $R_{b\tau}-A_{b\tau}/m_{5}$, $R_{b\tau}-\alpha$ and $R_{b\tau}-m_{3/2}$ planes. The color coding is the same as in Figure \ref{fig1}.}
\label{fig2}
\end{figure}

\begin{figure}[h!]
\centering
\subfigure{\includegraphics[scale=1.1]{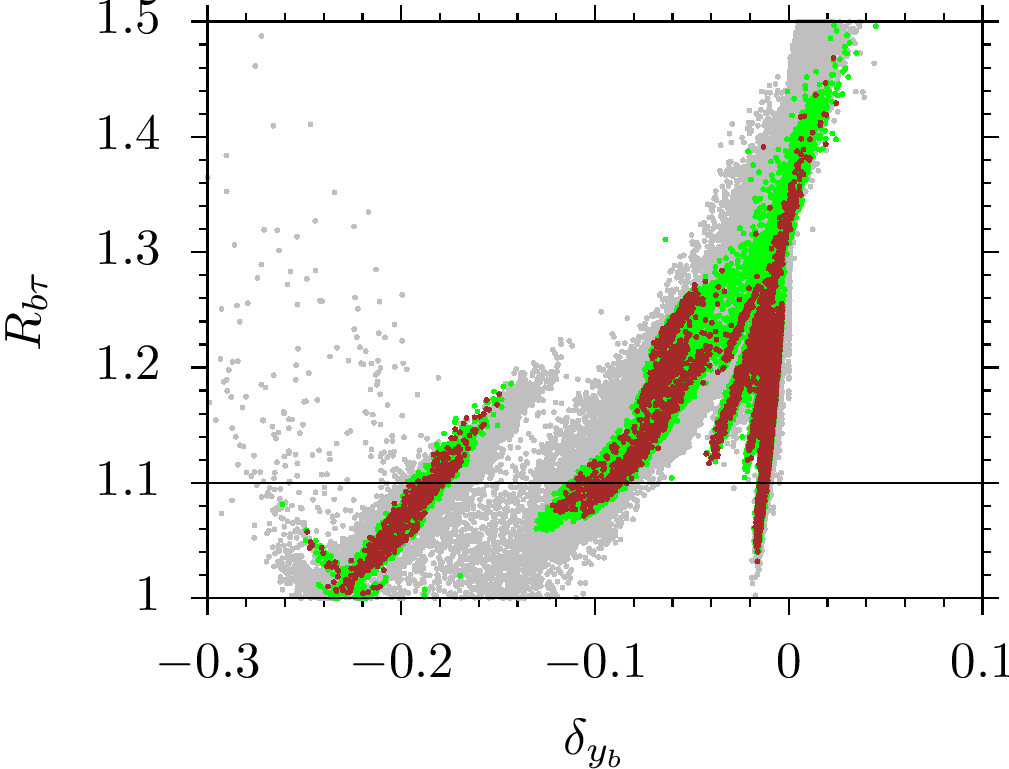}}
\subfigure{\includegraphics[scale=1.1]{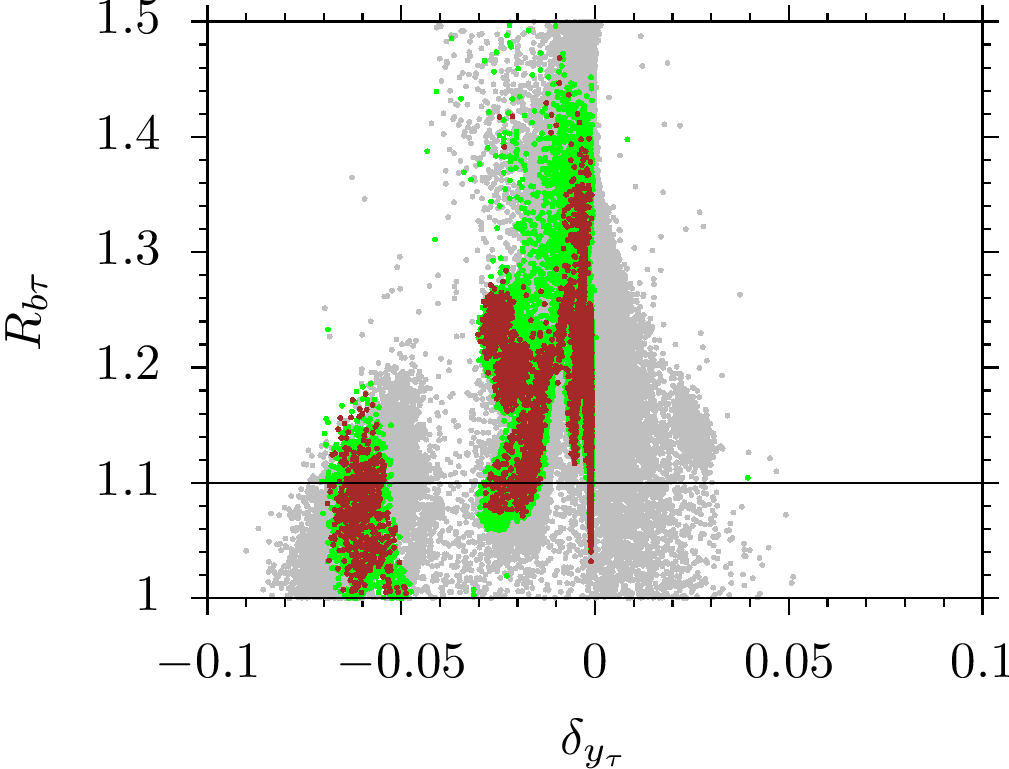}}
\subfigure{\includegraphics[scale=1.1]{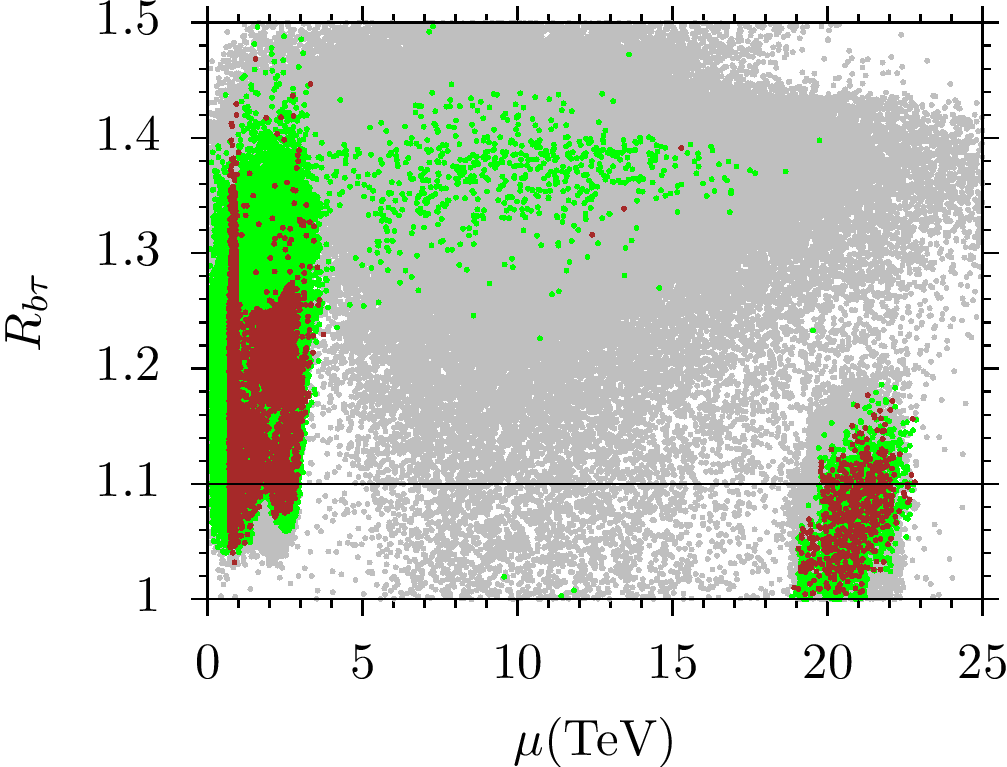}}
\subfigure{\includegraphics[scale=1.1]{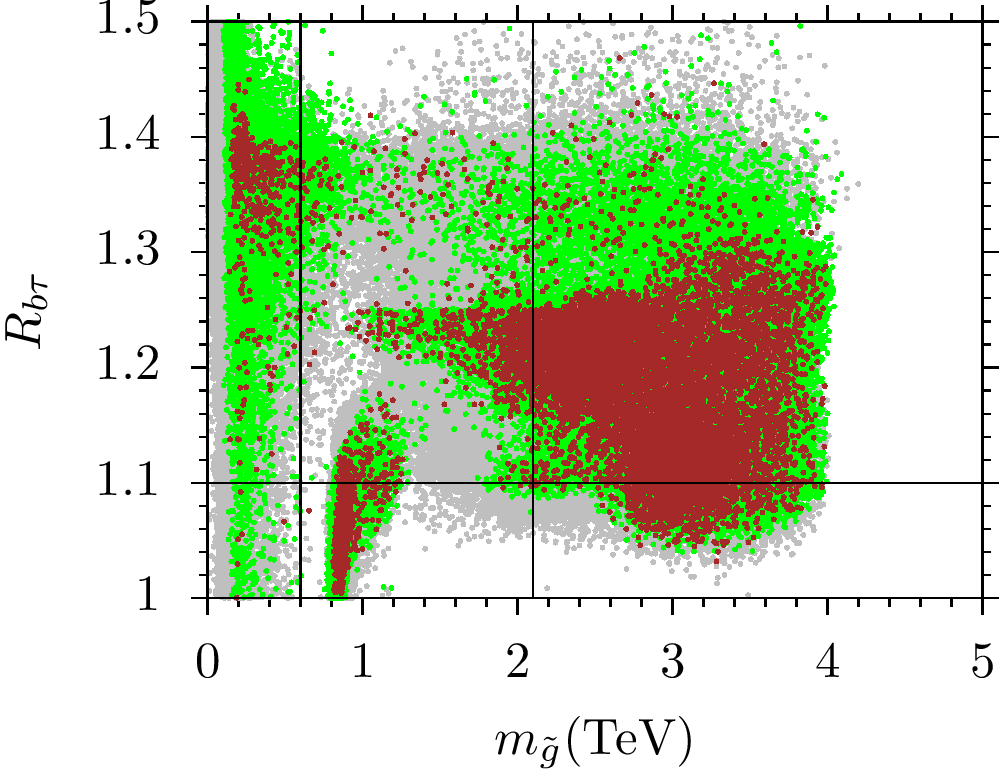}}
\caption{Plots in the $R_{b\tau}-\delta_{y_{b}}$, $R_{b\tau}-\delta_{y_{\tau}}$, $R_{b\tau}-\mu$ and $R_{b\tau}-m_{\tilde{g}}$ planes. The color coding is the same as in Figure \ref{fig1}.}
\label{fig3}
\end{figure}
%%%%%%%%%%%%%%%%%%%%%%%%%%%%%%
\begin{figure}[h!]
\centering
%\subfigure{\includegraphics[scale=1.1]{mirage_SU5_QYURbtau_mu.png}}
%\subfigure{\includegraphics[scale=1.1]{mirage_SU5_QYURbtau_glu.png}}
\subfigure{\includegraphics[scale=1.1]{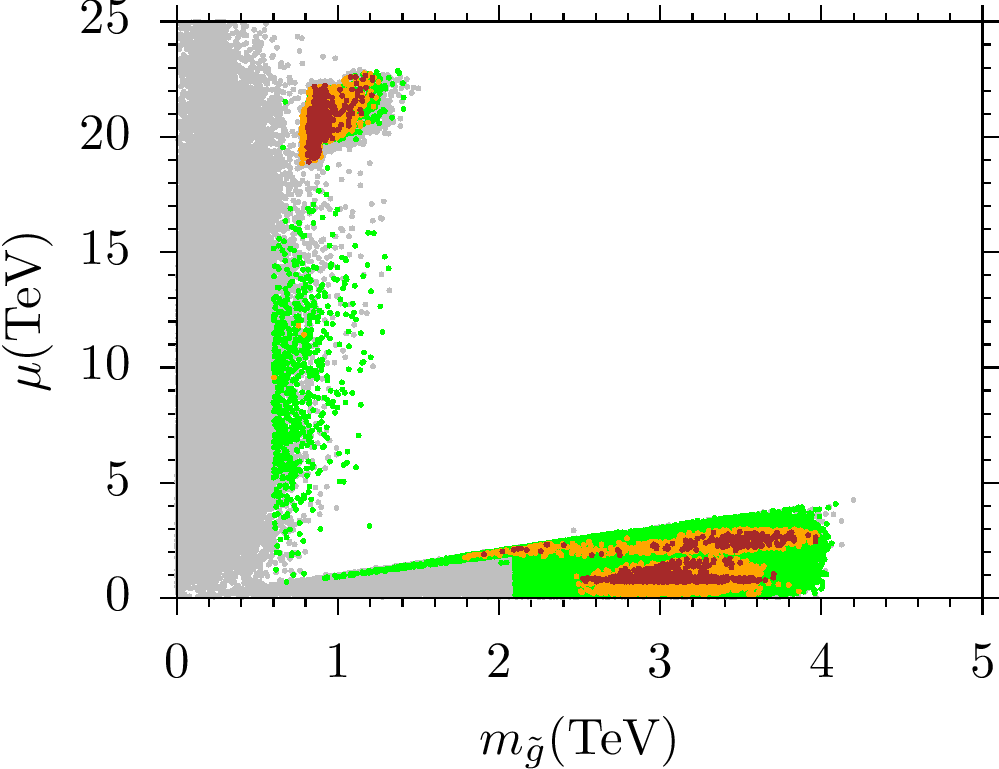}}
\subfigure{\includegraphics[scale=1.1]{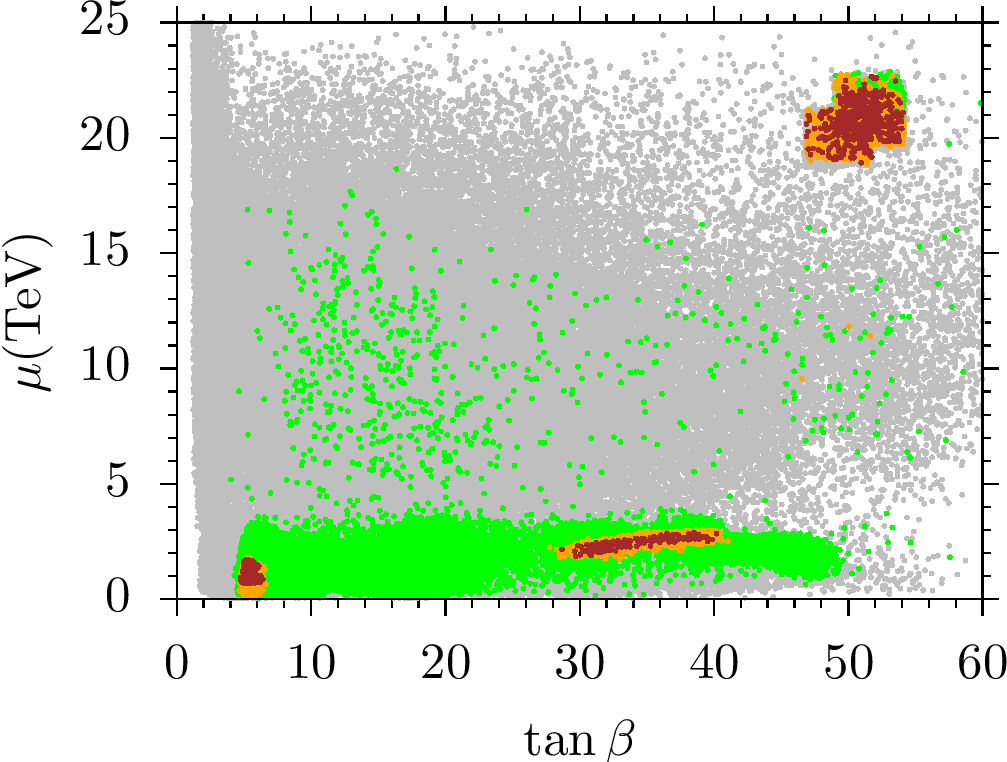}}

\caption{Plots in the  $\mu-m_{\tilde g}$ and $\mu-\tan\beta$  planes.  All points are consistent with the REWSB and LSP neutralino conditions. Green points satisfy the mass bounds and the constraints from rare $B-$meson decays. Orange points form a subset of green and the are compatible with the $b-\tau$ YU condition. Brown points, as a subset of orange, are allowed within $5\sigma$ by the WMAP bound on the relic abundance of LSP neutralino.}
\label{fig4}
\end{figure}

We start with the fundamental parameter space of $b-\tau$ YU, and discuss how it shapes the low energy implications. We display in figure the solutions compatible with $b-\tau$ YU in the $R_{b\tau}-m_{10}$, $R_{b\tau}-m_{5}$, $R_{b\tau}-M_{1/2}$, and $R_{b\tau}-\tan\beta$ planes. All points are consistent with the REWSB and LSP neutralino LSP conditions. Green points satisfy the mass bounds and the constraints from  rare $B-$meson decays. Brown points are a subset of green points, and they are compatible with the WMAP bound on the relic density of the LSP neutralino within $5\sigma$. The region below the horizontal line at $R_{b\tau}=1.1$ represents $b-\tau$ YU to within $10\%$. We see from the $R_{b\tau}-m_{10}$ plane that $b-\tau$ YU is mostly realized for $m_{10}\lesssim 9$ TeV. However, a second region with $m_{10}\sim 15-18$ TeV yields solutions with even more precise $b-\tau$ YU. We find a similar pattern in $R_{b\tau}-m_{5}$ plane where we find solutions with $10\%$ or better YU  for  $m_{5}\lesssim 3$ TeV and $m_{5}\gtrsim 12$ TeV. Comparing with the $R_{b\tau}-M_{1/2}$ plane shown in the lower left panel, we find that solutions with large values of $m_{10}$ and $m_{5}$ tend to have $M_{1/2} \lesssim$ 1 TeV, while smaller $m_{10}$ and $m_{5}$ values require rather large values of $M_{1/2} (\gtrsim 3$ TeV). In addition, we can identify two different regions for $b-\tau$ YU, as seen from the $R_{b\tau}-\tan\beta$ plane, one of which has $\tan\beta \lesssim$ 10, while the second one corresponds to $\tan\beta\gtrsim 30$. This looks similar to the results shown in \cite{Baer:2012by}; however, no solution for $\tan\beta \lesssim 20$ was in \cite{Baer:2012by} allowed by the Higgs boson mass constraint.

We continue the discussion with plots in the $R_{b\tau}-A_{t}/m_{10}$, $R_{b\tau}-A_{b\tau}/m_{5}$, $R_{b\tau}-\alpha$ and $R_{b\tau}-m_{3/2}$ planes with the color coding as in Figure \ref{fig1}. The $b-\tau$ YU condition requires negative $A_{t}$ ($\lesssim -1.5$), while $A_{b\tau}$ is found to be always positive and large. This opposite behavior of the trilinear couplings arises from the threshold corrections to b-quark Yukawa coupling and the Higgs boson mass bound. As mentioned before, $b-\tau$ YU requires negative large threshold corrections to $y_{b}$, and according to Eq.(\ref{eq:deltab}), such contributions can be obtained if $A_{t}$ is negative and its magnitude is large enough to cancel the positive contributions from the term containing $\mu$ and gluino mass ($m_{\tilde{g}}$). On the other hand, the Higgs boson mass constraint requires enhancements from large $\tan\beta$ (see also \cite{Baer:2012by}), as well as suitably large and positive $A_{b}$ and $A_{\tau}$ as can be seen from Eq.(\ref{eq:higgscor}). Here we want to make a comment that one can have $A_{t}>$ 0 if we take $\mu <$ 0 \cite{Gogoladze:2011db}.

 Finally we display the solutions in the $R_{b\tau}-\alpha$ and $R_{b\tau}-m_{3/2}$ planes. {There is an upper bound  $\alpha \lesssim 5$, because larger values than this significantly lowers $M_{3}$ at the GUT scale, which yields inconsistently light gluino masses in the low scale mass spectrum.} The results displayed in the $R_{b\tau}-m_{3/2}$ plane reveals no specific correlation between $b-\tau$ YU and $m_{3/2}$.

Before concluding this section, we discuss the impact of the threshold corrections on Yukawa coupling unification and their low energy implications. Figure \ref{fig3} displays the results with plots in the $R_{b\tau}-\delta_{y_{b}}$, $R_{b\tau}-\delta_{y_{\tau}}$, $R_{b\tau}-\mu$ and $R_{b\tau}-m_{\tilde{g}}$ planes. The color coding is the same as in Figure \ref{fig1}. As can be seen from the $R_{b\tau}-\delta_{y_{b}}$ plane, $b-\tau$ YU is controlled mostly by the contribution due to $y_{b}$, which are required to be from about $-0.24$ to $-0.2$ \cite{Gogoladze:2010fu}. Beyond this range, $y_{b}$ starts deviating from $b-\tau$ YU due to large corrections.  On the other hand, the results in the $R_{b\tau}-\delta_{y_{\tau}}$ plane show that the required contribution to $y_{\tau}$ is rather negligible as compared to $y_{b}$. 

The right amount of threshold contributions to $y_{b}$ also bounds the $\mu-$term, {and the solutions with low $\mu-$term ($\lesssim 3$ TeV) can be fit by $\delta_{y_{b}}$ given in Eq.(\ref{eq:deltab}) to yield $b-\tau$ YU solutions.} Such solutions might yield interesting implications at the low scale. For instance, since the fine-tuning is calculated in terms of $\mu$, such solutions can also be favored by a demand of lower fine-tuning. Besides, lower $\mu$ values can yield lighter Higgsinos, which may play an important role in DM phenomenology. These solutions also imply $m_{\tilde{g}} \gtrsim 2.1$ TeV, with $b-\tau$ YU becoming more precise for $m_{\tilde{g}} \gtrsim 3$ TeV. {However, these contributions also bound the gluino mass from above at about 4 TeV.} The mass scale for the gluino (after excluding the solutions with $m_{\tilde{g}} \leq 2.1$ TeV) just lies in the testable region in the near future collider experiments. The gluino is expected to be probed up to 3 TeV or so, when the $3000$ fb$^{-1}$ lumonosity is reached in the High Luminosity LHC (HL-LHC) experiments with 13-14 TeV center of mass energy \cite{Baer:2016wkz,Cakir:2014nba}. One can also expect its mass to be probed up to 4 TeV and beyond in the Future Circular Collider (FCC) experiments, conducted with 100 TeV center of mass energy.

Another region compatible with $b-\tau$ YU is realized for $\mu \gtrsim 19$ TeV, which is also discussed above. However, the gluinos in this case is as light as a TeV or so, which is allowed only if the gluino is nearly degenerate in mass with the LSP neutralino.
%SR%%%%%%%%%%%%5

{In Figure~\ref{fig4} we show plots in the $\mu-m_{\tilde g}$ and $\mu-\tan\beta$ planes.  All points are consistent with the REWSB and LSP neutralino conditions. Green points satisfy the mass bounds and the constraints from rare $B-$meson decays. Orange points form a subset of green and they are compatible with the $b-\tau$ YU condition. Brown points, as a subset of orange, are allowed within $5\sigma$ by the WMAP bound on the relic abundance of LSP neutralino. These plots will help us understand the differences between this study and the previous ones. 

In Ref.~\cite{Baer:2012cp}, $SO(10)$ boundary conditions were considered with $M_{1/2}$ and $m_{16}$ as the universal guagino and scalar mass parameters respectively. The requirement of $t-b-\tau$ YU yields NLSP gluinos (but the LSP neutralino relic density was not correct), and the rest of the spectrum was very heavy and hard to test in any future experiment. Recall that even in $t-b-\tau$ YU, the threshold corrections to $y_{b}$ are the most important ones \cite{Gogoladze:2010fu}. Also note that YU with $\mu >$ 0, requires relatively large values of $\mu$ . We can avoid this problem if we use $\mu <$ 0 \cite{Gogoladze:2010fu}. The compatiblity issue of YU and the NLSP gluino (with correct relic density) can be overcome if we consider non-universal guigions \cite{Raza:2014upa}. In this way one can obtain NLSP gluino with $t-b-\tau$ YU, and if  we relax to $b-\tau$ YU, we can find NLSP stop as well with the correct relic density. In Ref.~\cite{Baer:2012by}, there is scalar mass non-universality but the gaugino mass parameter is universal, i.e $M_{1/2}$. In this way the possiblity to obtain a gluino mass appropriate both for YU and as an NLSP was lost, and the gluino turns out to be heavy. With $\mu >$ 0, large values of $\mu$ are required for YU and therefore one could not have light charginos (so no chargino-neutralino coaanihilation). Moreover, it can also be understood why there is no $m_{A}$ resonance solutions since with $m_{A}^2=2|\mu|^{2}+ m_{H_{u}}^{2}+m_{H_{d}}^{2}$ \cite{Martin:1997ns}, the former turns out to be heavy. {{We also know that for YU, we usually have a heavy SUSY spectrum under these circumstances, and the stop should be the NLSP \cite{Gogoladze:2011be}, which was earlier reported in \cite{Baer:2012by}. }}

In this study, we have non-universal guaginos as well as a non-universal scalar sector. We have more freedom to achieve $b-\tau$ YU with a variety of NLSPs. Figure~\ref{fig4} shows the interplay of $\mu$, $\tan\beta$ and $m_{\tilde g}$. In the $\mu-m_{\tilde g}$ plot, we can clearly see that with large values of $\mu$ we require $m_{\tilde g} \sim$ 1 TeV (as mentioned above). On the other hand, for larger values of the gluino mass, $\mu$ can be light. This will open up new DM channels to achieve the correct relic density. The $\mu-\tan\beta$ plot also clearly shows that the solutions with low $\tan\beta$, which were previously ruled out because of a low Higgs mass  in \cite{Baer:2012by}, are now available because we have solutions with heavier gluinos that push up the stop mass which, in turn, uplifts the Higgs mass.}

%%%%%%%%%%%
\section{Sparticle Mass Spectrum}
\label{sec:mass}

\begin{figure}[h!]
\centering
\subfigure{\includegraphics[scale=1.1]{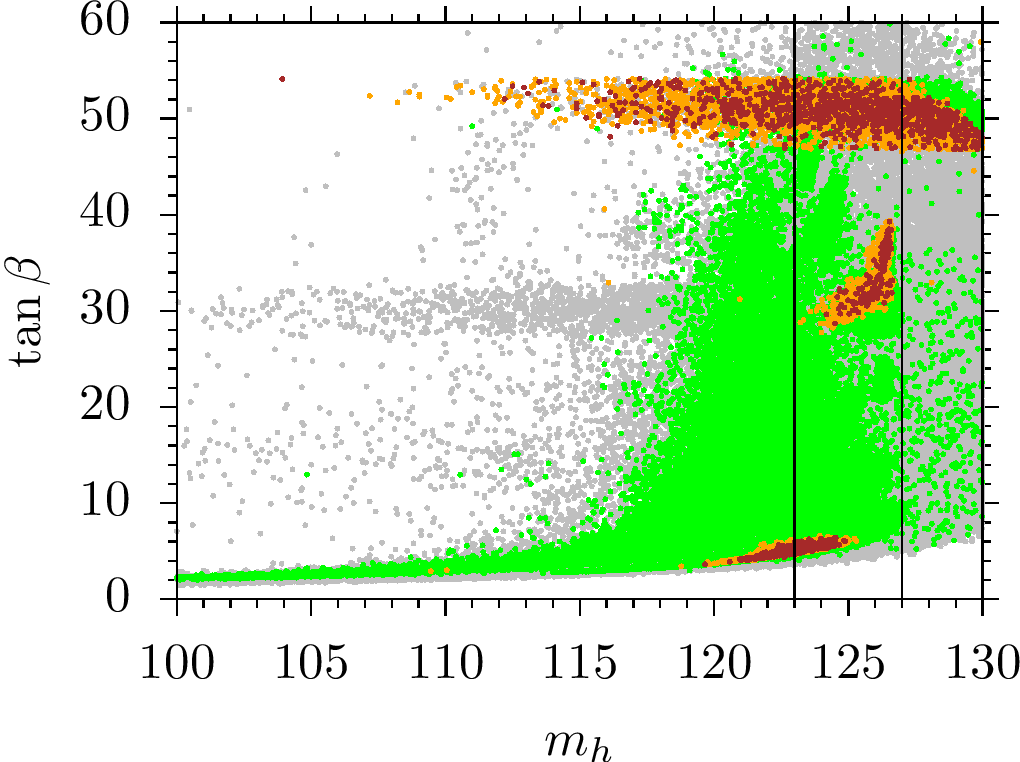}}
\subfigure{\includegraphics[scale=1.1]{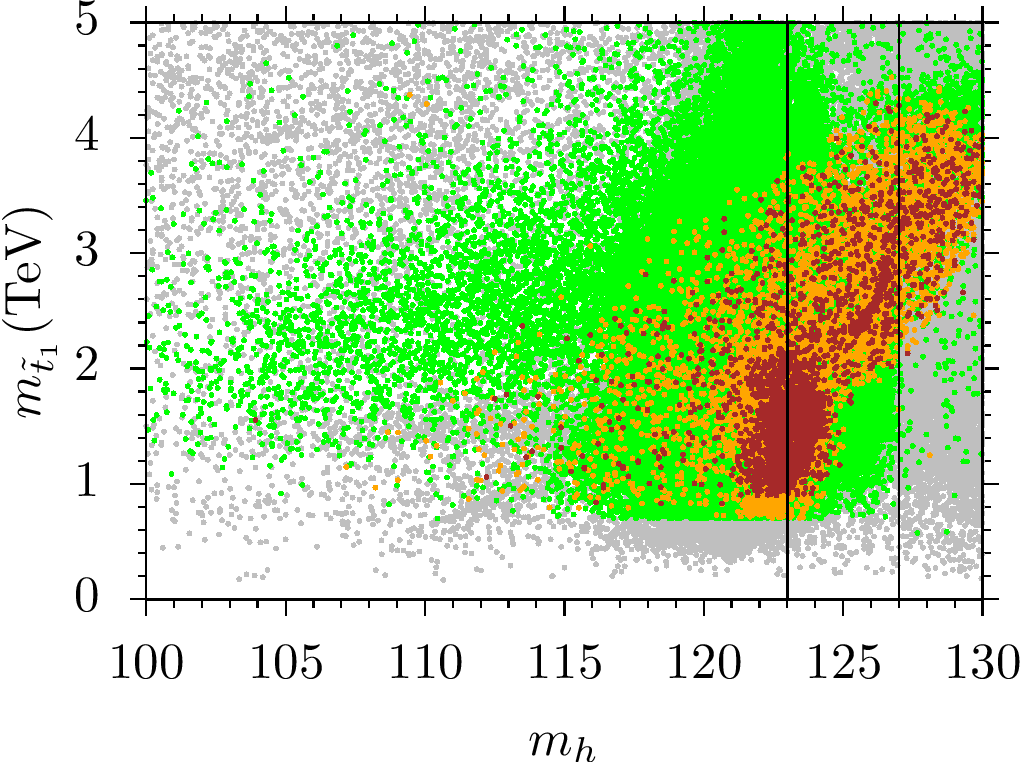}}
\caption{Plots in the $\tan\beta -m_{h}$ and $m_{\tilde{t}_{1}}-m_{h}$ planes. The color coding is the same as in Figure \ref{fig4}. The Higgs mass bound is not applied, but it is indicated with two vertical lines at $m_{h}=123$ GeV and $m_{h}=127$ GeV.}
\label{fig5}
\end{figure}

{As previously mentioned the parameter space consistent with $b-\tau$ YU can be restricted by the Higgs boson mass.  For example, solutions with low $\tan\beta$ compatible with $b-\tau$ YU can be ruled out since they yield a small Higgs mass \cite{Baer:2012by}}. Figure \ref{fig5} represents the results for the Higgs boson mass with plots in the $\tan\beta -m_{h}$ and $m_{\tilde{t}_{1}}-m_{h}$ planes. All points are consistent with the REWSB and LSP neutralino conditions. Green points satisfy the mass bounds {(we do not apply the Higgs mass bound here)} and the constraints from rare $B-$meson decays. Orange points form a subset of green points and they are compatible with $10\%$ or better $b-\tau$ YU condition. Brown points serves as a subset of orange points and represent solutions which are allowed by the WMAP bound on the relic abundance of LSP neutralino within $5\sigma$. The Higgs mass bound is not applied, but it is indicated with two vertical lines at $m_{h}=123$ GeV and $m_{h}=127$ GeV. We see from the $\tan\beta -m_{h}$ plane that the Higgs boson mass barely reaches 125 GeV under the $b-\tau$ YU condition for $\tan\beta \lesssim 10$, and $b-\tau$ YU cannot be realized beyond this region until $\tan\beta \gtrsim 30$. The reason is that the stops cannot contribute enough to the Higgs boson mass in this region, and the contributions from the sbottom and stau can only take part in the Higgs boson mass calculation for $\tan\beta \gtrsim 30$. Also, the stops cannot be heavier than about 2 TeV in the low $\tan\beta$ region as seen from the $m_{\tilde{t}_{1}}-m_{h}$ plane shown in the right panel.

\begin{figure}[ht!]
\centering
\subfigure{\includegraphics[scale=1.1]{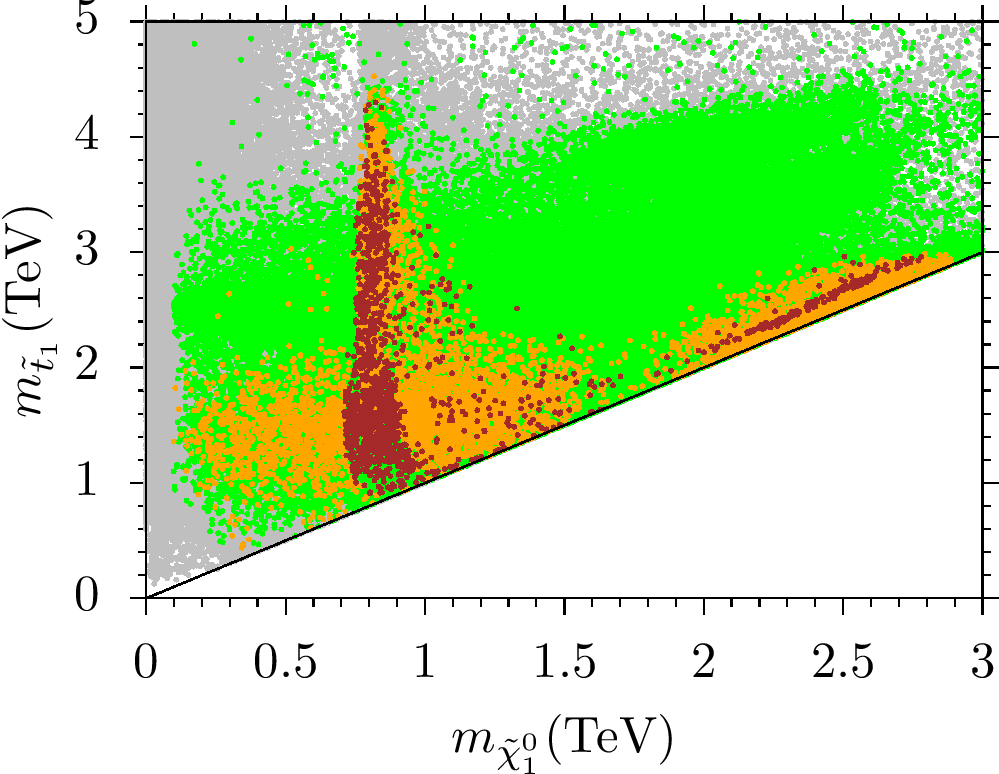}}
\subfigure{\includegraphics[scale=1.1]{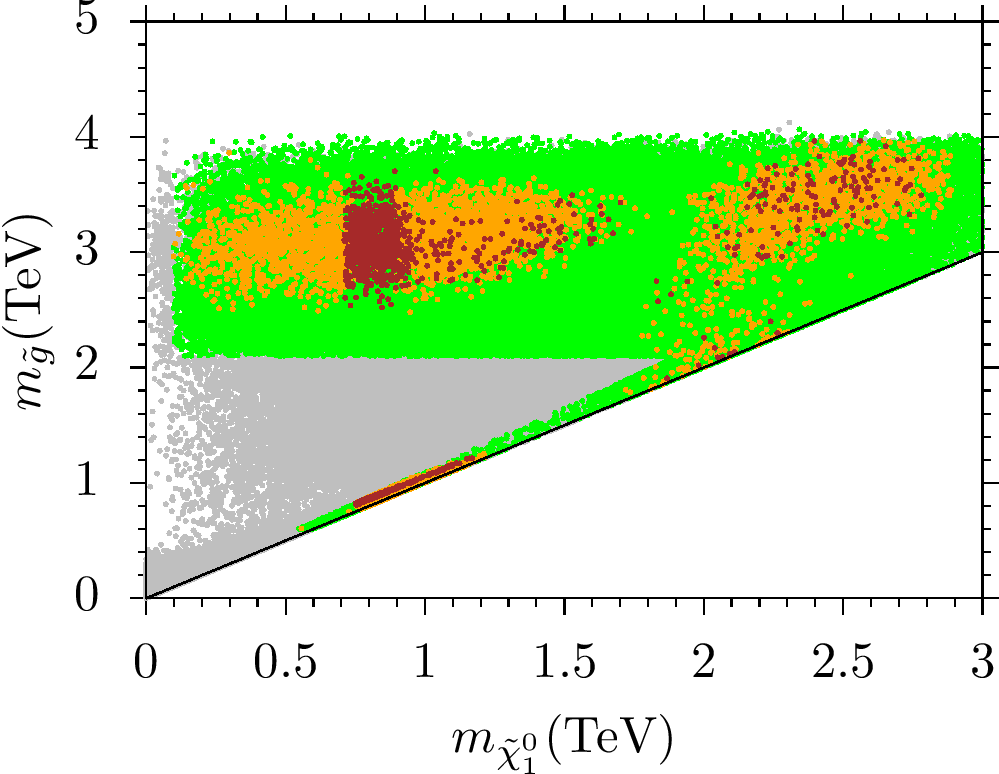}}
\subfigure{\includegraphics[scale=1.1]{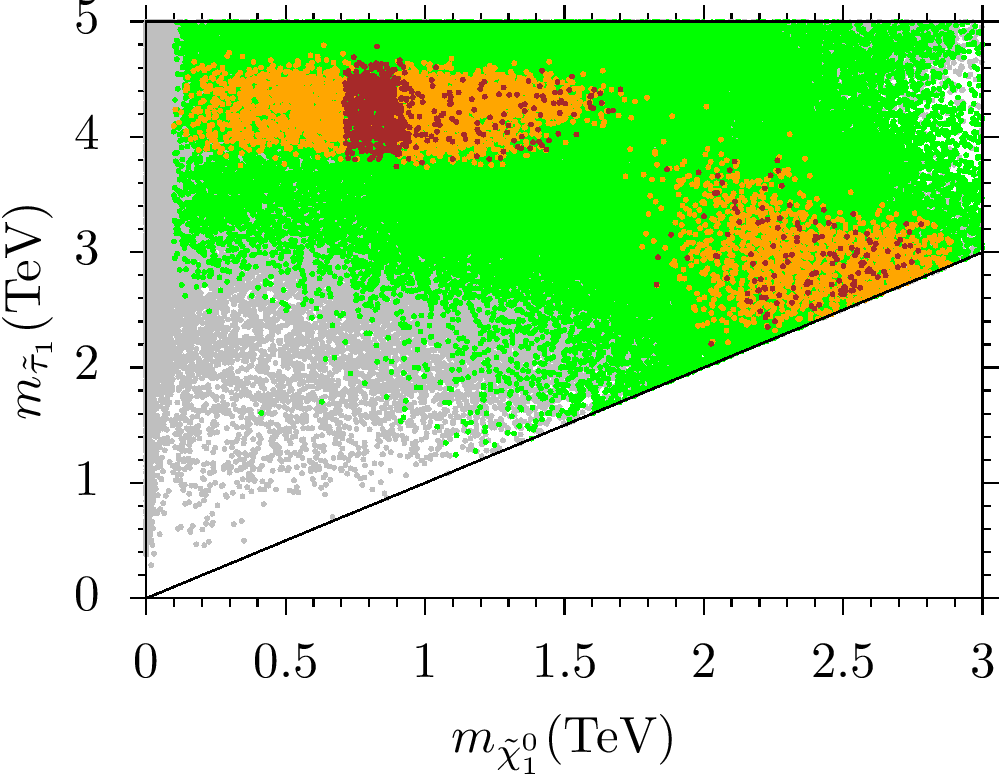}}
\subfigure{\includegraphics[scale=1.1]{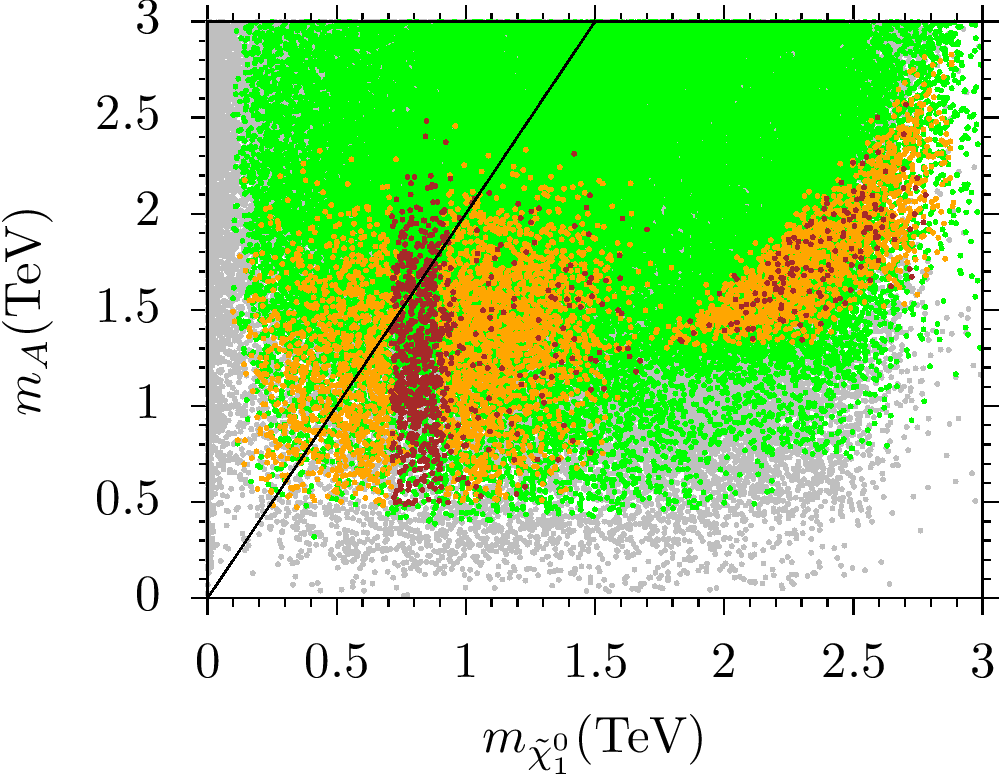}}
\caption{Plots in the $m_{\tilde{t}_{1}}-m_{\tilde{\chi}_{1}^{0}}$, $m_{\tilde{g}}-m_{\tilde{\chi}_{1}^{0}}$, $m_{\tilde{\tau}_{1}}-m_{\tilde{\chi}_{1}^{0}}$ and $m_{A}-m_{\tilde{\chi}_{1}^{0}}$. The color coding is the same as in Figure \ref{fig4}. In addition, the Higgs boson mass bound is also applied. The diagonal lines in the first three plots indicate the mass degeneracy between the particles, while in the $m_{A}-m_{\tilde{\chi}_{1}^{0}}$ plane, the diagonal line indicates the regions where $m_{A}=2 m_{\tilde{\chi}_{1}^{0}}$.}
\label{fig6}
\end{figure}

As discussed above, we expect a variety of NLSPs in this study. Figure \ref{fig6} displays the SUSY spectrum with plots in the $m_{\tilde{t}_{1}}-m_{\tilde{\chi}_{1}^{0}}$, $m_{\tilde{g}}-m_{\tilde{\chi}_{1}^{0}}$, $m_{\tilde{\tau}_{1}}-m_{\tilde{\chi}_{1}^{0}}$ and $m_{A}-m_{\tilde{\chi}_{1}^{0}}$ planes. The color coding is the same as in Figure \ref{fig4}. The diagonal lines in the first three plots indicate the mass degeneracy between the listed particles, and in the $m_{A}-m_{\tilde{\chi}_{1}^{0}}$ plane, the black line indicates the region along which $m_{A}=2 m_{\tilde{\chi}_{1}^{0}}$. The $m_{\tilde{t}_{1}}-m_{\tilde{\chi}_{1}^{0}}$ plane shows that the stop can only be as heavy as about 4.5 TeV, and it is nearly degenerate in mass with the LSP neutralino mass from about 800 GeV to 1.7 TeV. Such solutions represent the stop-neutralino coannihilation scenario, which reduce the relic abundance of LSP neutralino to the desired value. As we previously showed, the gluino mass is bounded from above at about 4 TeV, and its mass is nearly degenerate with the LSP neutralino from about 800 GeV to 1.2 TeV. By requiring $m_{\tilde{g}}\lesssim 1.1 \, m_{\tilde{\chi}_{1}^{0}}$ in this region, the current gluino mass bound excludes most of these solutions. 

It is also possible to realize mass degeneracy between the NLSP gluino and LSP neutralino between about $1.8-2.5$ TeV, as seen from the $m_{\tilde{g}}-m_{\tilde{\chi}_{1}^{0}}$ plot. {These gluino-neutralino coannihilation processes also satisfy the WMAP relic density bounds. In addition to stop and gluino, we also have available the stau-neutralino coannihilation channel  to achieve the desired relic density. We see that the mass degeneracy between the NLSP stau and the lightest neutralino occurs for $m_{\tilde{\tau}_{1}}\approx m_{\tilde{\chi}_{1}^{0}} \gtrsim 2$ TeV}.

Besides the SUSY particles, the CP-odd Higgs boson can take part in reducing the relic LSP neutralino abundance down to the allowed range determined by the current WMAP measurements. The $m_{A}-m_{\tilde{\chi}_{1}^{0}}$ plane shows that it is possible to find solutions with $m_{A} \approx 2 m_{\tilde{\chi}_{1}^{0}}$, for $m_{A}\sim 1.5$ TeV. These solutions allow processes in which two LSP neutralinos annihilate into a CP-odd Higgs boson.

\begin{figure}[h!]
\centering
\subfigure{\includegraphics[scale=1.1]{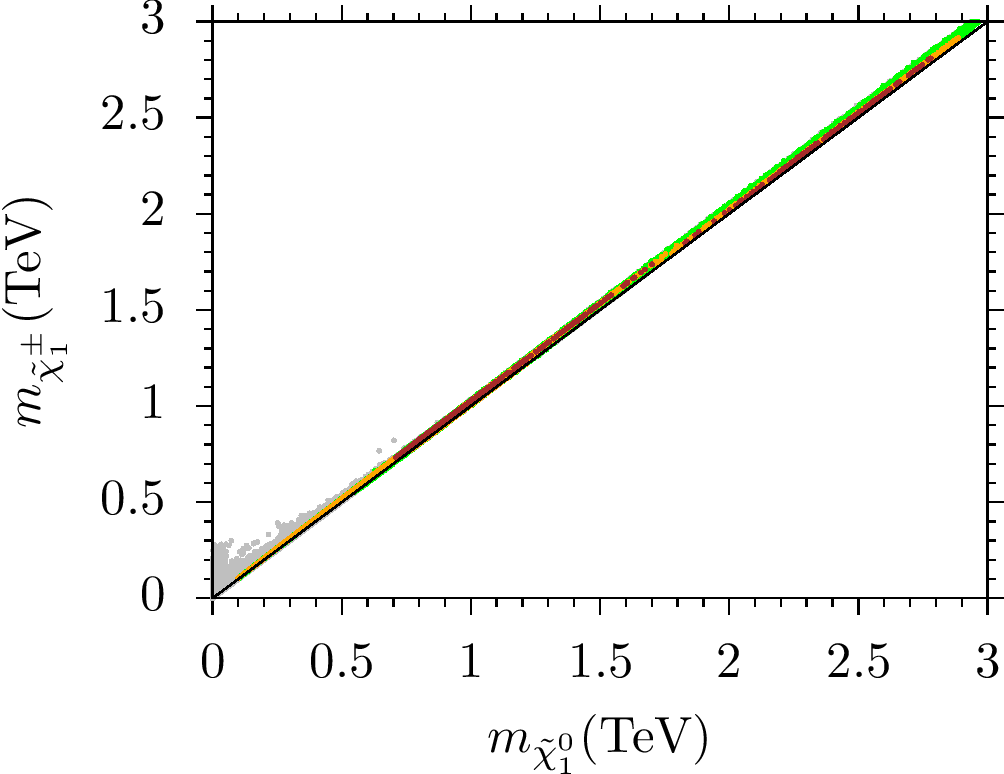}}
\subfigure{\includegraphics[scale=1.1]{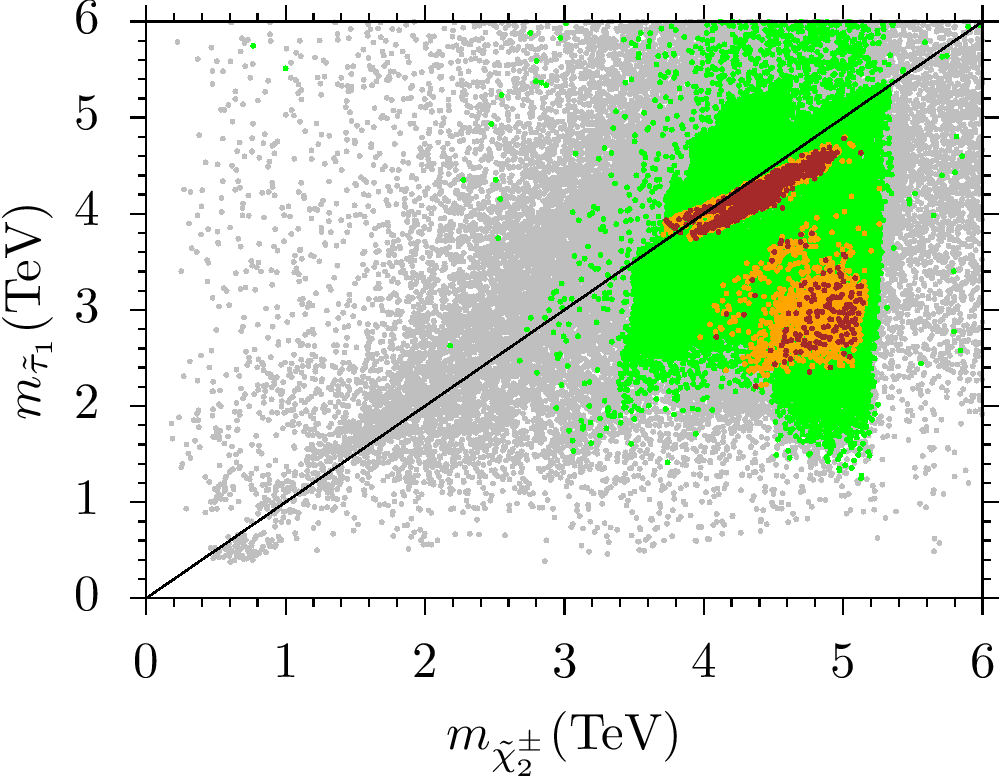}}
\caption{Plots in the $m_{\tilde{\chi}_{1}^{\pm}}-m_{\tilde{\chi}_{1}^{0}}$ and $m_{\tilde{\tau}_{1}}-m_{\tilde{\chi}_{2}^{\pm}}$ planes. The color coding is the same as Figure \ref{fig4}. The diagonal lines indicate the regions predicting mass degeneracy between the particles.}
\label{fig7}
\end{figure}

We present masses of the SUSY fermions in Figure~\ref{fig7} with plots in the $m_{\tilde{\chi}_{1}^{\pm}}-m_{\tilde{\chi}_{1}^{0}}$ and $m_{\tilde{\tau}_{1}}-m_{\tilde{\chi}_{2}^{\pm}}$ planes. The color coding is the same as in Figure \ref{fig4}. The diagonal lines indicate the region of mass degeneracy between the relevant particles. The $m_{\tilde{\chi}_{1}^{\pm}}-m_{\tilde{\chi}_{1}^{0}}$ plane shows  the lightest chargino that is degenerate in mass with the LSP neutralino consistent with $b-\tau$ YU. Their masses lie in the range from about 0.7 TeV to 2.8 TeV. Such a mass degeneracy favours coannihilation processes involving the LSP neutralino and the lightest chargino for reducing the relic abundance of the LSP neutralino. Degenerate chargino and neutralino are characteristic of the solutions leading to either wino-like or Higgsino-like LSP neutralino. We also plot the results for the second chargino mass versus the stau mass, and we find $m_{\tilde{\chi}_{2}^{\pm}} \gtrsim 4$ TeV.

The recent LHC analyses have revealed new mass bounds on the charginos based on their decay modes. If the chargino decays into stau along with a suitable SM particle, then the solutions with $m_{\tilde{\chi}_{i}^{\pm}}\lesssim 1.1$ TeV are excluded \cite{CMS:2017fdz}, where $i=1,2$ stand for the lighter and heavier chargino respectively. On the other hand, if the staus are too heavy the charginos cannot decay, and the mass bound becomes $m_{\tilde{\chi}_{i}^{\pm}} \gtrsim 500$ GeV \cite{Sirunyan:2017zss}. Since the lightest chargino is degenerate in mass with the LPS neutralino, it cannot decay into staus, and the mass bound $m_{\tilde{\chi}_{1}^{\pm}} \gtrsim 500$ GeV holds for these solutions. In our results $b-\tau$ YU compatible with all the constraints listed in Section \ref{sec:scan} bounds the chargino mass $m_{\tilde{\chi}_{1}^{\pm}}\gtrsim 700$ GeV, and these solutions are expected to be tested quite soon. On the other hand, since $m_{\tilde{\chi}_{2}^{\pm}} \geq m_{\tilde{\tau}_{1}}$, the second chargino is allowed to decay into stau over most of the parameter space; however its mass is bounded from below, namely $m_{\tilde{\chi}_{2}^{\pm}} \gtrsim 4$ TeV, which is well beyond the reach of current experiments.

\section{Dark Matter Implications}
\label{sec:DM}

\begin{figure}[h!]
\centering
\subfigure{\includegraphics[scale=1.1]{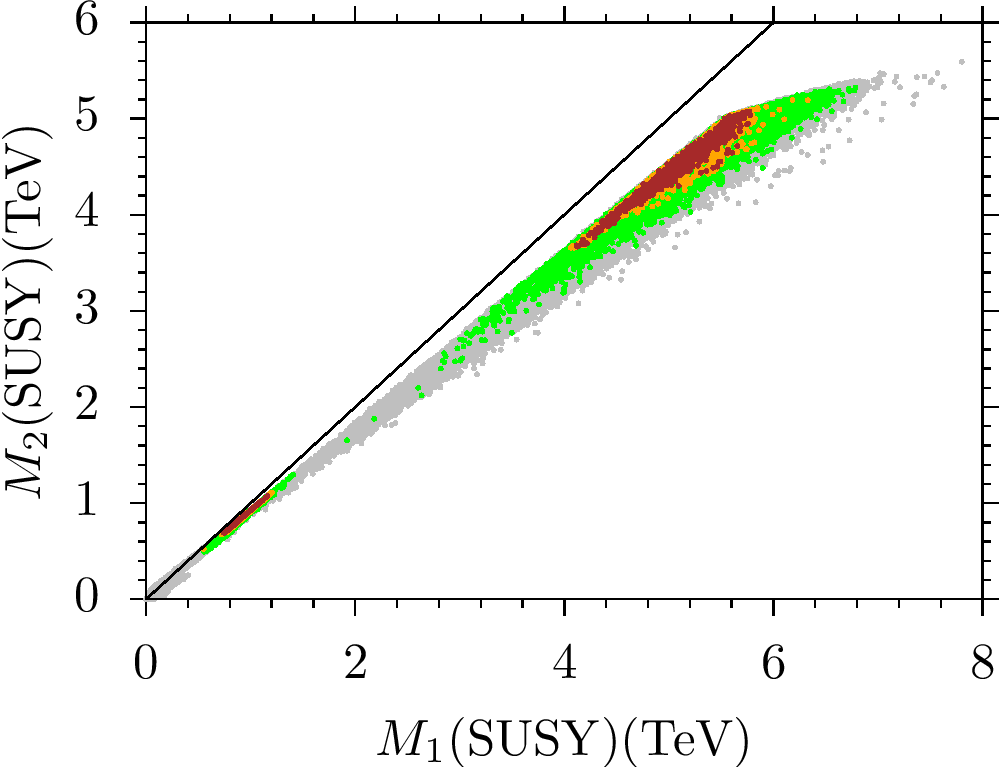}}
\subfigure{\includegraphics[scale=1.1]{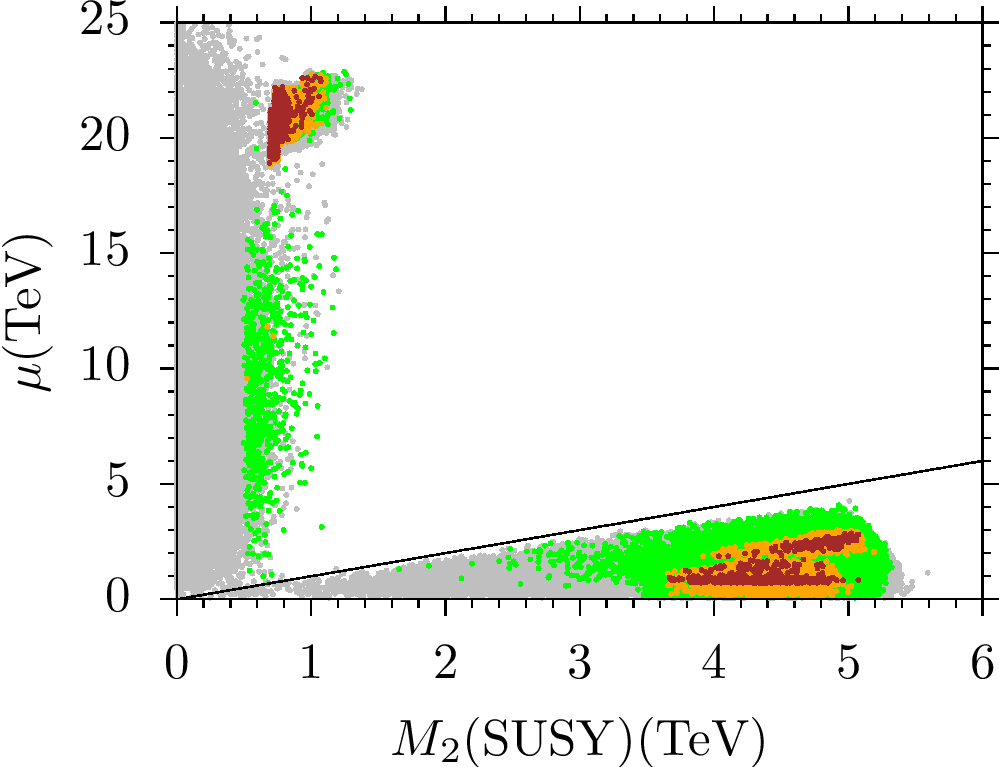}}
\caption{Plots in the $M_{2}-M_{1}$ and $\mu - M_{2}$ planes. The color coding is the same as in Figure \ref{fig4}. The black lines indicate the regions where the plotted parameters are equal to each other.}
\label{fig8}
\end{figure}

Finally, in this section we study DM in our model with $b-\tau$ YU in the presence of mirage mediation and discuss the implications for the current and near future experiments. In the MSSM framework the bino, wino, and two Higgsinos mix with each other and form the four neutralino mass eigenstates. In this context, the LSP neutralino can be mostly one  or a mixture of them. This yields a variety of phenomenology with implications, which can be tested by the on going and future experiments. Figure \ref{fig8} shows the plots at low scale in the $M_{2}-M_{1}$ and $\mu - M_{2}$ planes. The color coding is the same as in Figure \ref{fig4}. The black lines indicate the regions where the plotted parameters are equal to each other. $M_{1}$, $M_{2}$ and $\mu$ represent the masses of bino, wino and Higgsinos respectively. We see in the $M_{2}-M_{1}$ plot that the bino is always heavier than wino, even though some solutions can yield the LSP neutralino as a mixture of bino-wino for $M_{2}\sim M_{1} \lesssim 1.2$ TeV. These solutions can be identified with the gluino-LSP neutralino mass degeneracy as discussed in Section \ref{sec:btauYU}. On the other hand, despite masses of the same order in the second region with $M_{2}\gtrsim 4$ TeV (in brown), the LSP neutralino is not composed of either the bino or the wino. These neutralino eigenstates are instead composed of the third and fourth neutralino mass eigenstates. This region yields Higgsino-like LSP neutralino, which can be seen easily from the $\mu - M_{2}$ plane. For $M_{2}\gtrsim 4$ TeV, the Higgsino mass parameter $\mu$ is less than about 3 TeV for all the solutions. One can also see the gluino-LSP neutralino degeneracy solutions as a separate second region in which $\mu \gtrsim 20$ TeV.

If one assume the LSP neutralino as a DM candidate, the Higgsino-like LSP neutralino yields quite compelling results for the direct detection experiments. In these cases, since the scattering of DM on the nuclei proceeds through the Yukawa interactions, the scattering cross-section is expected to be high and either testable or excluded by the current constraints from the direct detection experiments. We show our results for the spin-independent (left) and spin-dependent (right) scattering cross-sections as a function of the LSP neutralino mass. In the $\sigma_{SI}-m_{\tilde{\chi}_{1}^{0}}$ plane, the dashed (solid) blue line represents the current (future) results from the SuperCDMS experiment \cite{Brink:2005ej}. The dashed (solid) line indicates the current (future) results from the LUX-Zeplin experiment \cite{Akerib:2018lyp}. In the $\sigma_{SD}-m_{\tilde{\chi}_{1}^{0}}$ plane, the solid black line represents the currrent bound from  Super-K \cite{Tanaka:2011uf}, while the orange solid line is set by the LUX results \cite{Akerib:2016lao}. The purple line is obtained from the collider analyses \cite{Khachatryan:2014rra}.  The dashed (solid) blue line shows the current (future) results from IceCube DeepCore.

\begin{figure}[h!]
\centering
\subfigure{\includegraphics[scale=1.2]{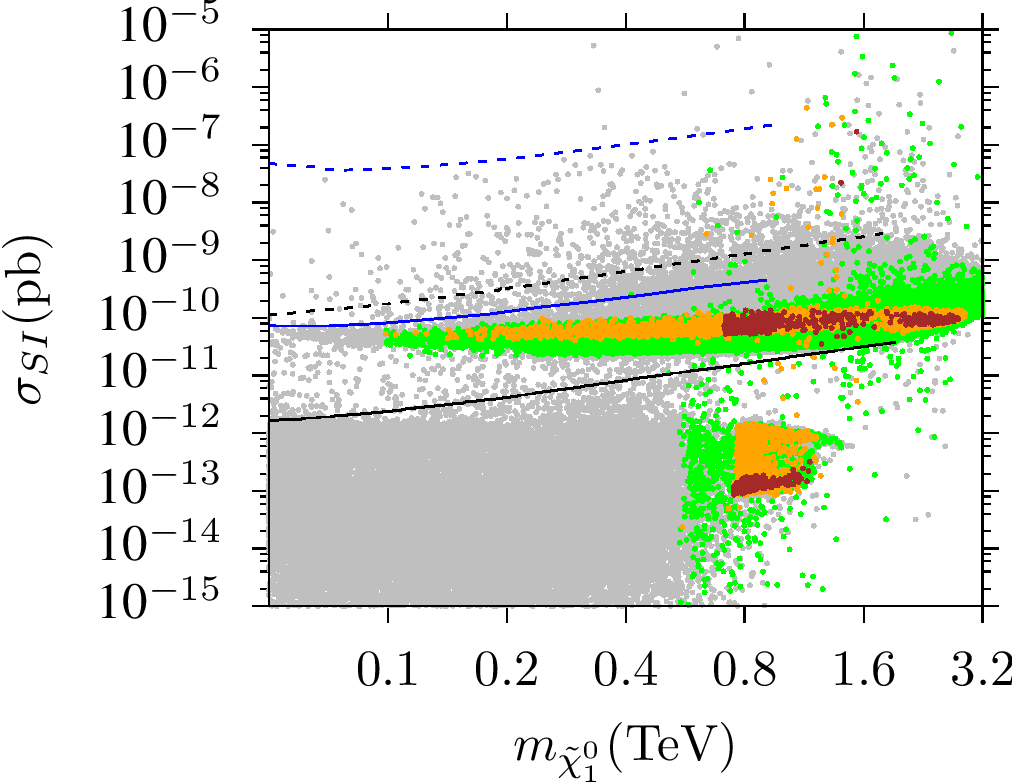}}%
\subfigure{\includegraphics[scale=1.2]{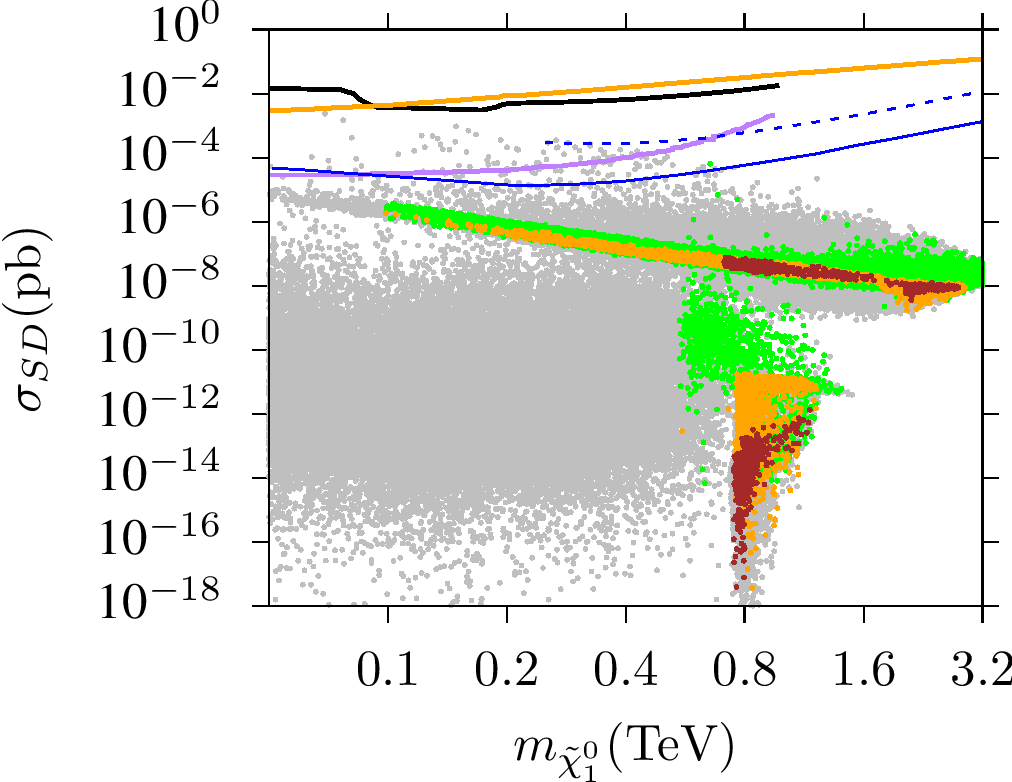}}
\caption{Spin-independent (left) and spin-dependent (right) scattering cross-sections versus the LSP neutralino mass. In the $\sigma_{SI}-m_{\tilde{\chi}_{1}^{0}}$ plane, the dashed (solid) blue line represents the current (future) results from the SuperCDMS experiment \cite{Brink:2005ej}. The dashed (solid) line indicates the current (future) results from the LUX-Zeplin experiment \cite{Akerib:2018lyp}. In the $\sigma_{SD}-m_{\tilde{\chi}_{1}^{0}}$ plane, the solid black line represents the currrent bound from Super-K \cite{Tanaka:2011uf}, while the orange solid line is set by the LUX results \cite{Akerib:2016lao}. The purple line is obtained from the collider analyses \cite{Khachatryan:2014rra}.  The dashed (solid) blue line shows the current (future) results from IceCube DeepCore.}
\label{fig9}
\end{figure}

As seen from the $\sigma_{SI}-m_{\tilde{\chi}_{1}^{0}}$ plane, the solutions yield mostly large scattering cross-sections, with some solutions accumulate in a region which yields relatively low scattering cross-section. The region with the higher scattering cross-sections is identified with Higgsino-like DM, while the other one yields bino-wino mixed DM. In the latter case, since the scattering happens through the gauge interactions between the bino-wino and quarks in nuclei, the DM scattering is considerably weaker. All solutions are allowed by the current and future bounds from the SuperCDMS experiment (dashed and solid blue lines), but the LUX-Zeplin (LZ) experiment could have a strong impact. Since all solutions lie below the dashed black line, they are compatible with the current LUX results. However, the projected exclusion (solid black) line is expected to provide more sensitive results such that all of the Higgsino-like DM solutions can be  tested in the near future.  The plot in the $\sigma_{SD}-m_{\tilde{\chi}_{1}^{0}}$ plane shows that our solutions are consistent with the current and future reaches of the experiments.

\begin{table}[h!]
\centering
\scalebox{0.8}{
\begin{tabular}{|l|c|c|c|c|c|}
\hline
\hline
                 &  Point 1 & Point 2  &  Point 3  & Point 4 & Point 5   \\
\hline
$m_{10}$        &  17810   & 6421   & 3338    & 3844 & 3989     \\
$m_{5}$        &  14020   & 2433   & 17210    & 4928 & 4504      \\
$\alpha$ &  2.52   & 2.58   & 3.68    & 2.67 & 2.66      \\
$m_{3/2}$ &  94680   & 33150   & 85290    & 81450 & 85930      \\
$M_{1}$         &  2371 & 9938 & 10379   & 9374  & 9058      \\
$M_{2} $          & 1194    & 4885    & 5222   & 4599   & 4456     \\
$M_{3}$         &  352.9  & 1276   & 1538   & 1189   & 1168  \\
$A_{t}/m_{10}$  &  -2.6  & -1.72  & -1.49   & -0.93  & -0.9   \\
$A_{b\tau}/m_{5}$  &  0.15  & 9.28  & 1.84   & 8.37  & 8.1   \\
$\tan\beta$       & 52.1    & 5.6    &  31  &  20.4   & 19.05  \\
$m_{H_d}$         & 11880   & 1001   & 5808   &  6685  & 6365   \\
$m_{H_u}$          & 3045    & 8131    & 909.1   & 3360  & 3678 \\
\hline
$m_h$                & {\bf 126.3}   & {\bf 123.9}    & {\bf 125.2}   & {\bf 125.1}  & {\bf 124.8} \\
$m_H$                 & 18487 & 944.8  & 1436  & 1707 & 1725   \\
$m_A$                 & 18367 & 938  & 1427  & 1696  & 1713       \\
$m_{H^{\pm}}$         & 18488 & 946.9  & 1439  & 1708 & 1726  \\
\hline
$m_{\tilde{\chi}^0_{1,2}}$
                &  {\color{red}1161},{\color{red}1174} & {\color{red}1047}, {\color{red} 1049} & {\color{red} 2026}, {\color{red} 2028}  & {\color{red} 811}, {\color{red} 813} & {\color{red} 774}, {\color{red} 777}   \\
$m_{\tilde{\chi}^0_{3,4}}$
               & 21042, 21042 & 4120, 4569 & 4395, 4796  & 3819, 4263  & 3699, 4117   \\

$m_{\tilde{\chi}^{\pm}_{1,2}}$
               &  {\color{red}1173}, 21083 & {\color{red} 1076}, 4108& {\color{red} 2042}, 4372  & {\color{red} 826}, 3794  & {\color{red} 789}, 3674   \\
\hline
$m_{\tilde{g}}$ &  {\color{red}1215}   & 2930  & 3469  & 2691 & 2657           \\
$m_{ \tilde{u}_{L,R}}$
               & 17850, 17729& 7403, 7356  & 5124, 4651  & 5244, 4810  & 5288, 4901     \\
$m_{\tilde{t}_{1,2}}$
               & 2699, 10443 & {\color{red} 1103}, 4923 & 2185, 2608 & 2262, 2795  & 2564, 2858     \\
\hline $m_{ \tilde{d}_{L,R}}$
                &17850, 14033 & 7403, 3295 & 5125, 17601  & 5245, 5531 & 5289, 5119      \\
$m_{\tilde{b}_{1,2}}$
                 & 9344, 10244 & 1318, 4981 & 2350, 16707  & 1678, 2470  & 1042, 2658   \\
\hline
$m_{\tilde{\nu}_{e,\mu}}$
              & 13924   & 4460 & 17636 & 5886 & 5486       \\
$m_{\tilde{\nu}_{\tau}}$   & 11360 & 4126  & 17322  & 4847 & 4506    \\
\hline
$m_{ \tilde{e}_{L,R}}$
              & 13906, 18005 & 4455, 7164 & 17636, 5124  & 5888, 5303  & 5486, 5309    \\
$m_{\tilde{\tau}_{1,2}}$
              &  11336, 14059 & 4127, 6723 & {\color{red} 2206}, 17271  & 2415, 4810  & 2924, 4476    \\
\hline

$\sigma_{SI}({\rm pb})$
             &  $0.22\times 10^{-12}$& 0.97$\times 10^{-10}$ & 1.17$\times 10^{-10}$ & 0.68$\times 10^{-10}$ & $0.72\times 10^{-10}$ \\

$\sigma_{SD}({\rm pb})$
             &  0.56$\times 10^{-12}$ &0.36$\times 10^{-7}$ &1.18$\times 10^{-8}$   &0.79$\times 10^{-7}$  & $0.96 \times 10^{-7}$\\
$\Omega_{CDM}h^{2}$
            &  0.116 & 0.125 &0.12  & 0.123  & 0.118 \\
\hline
$R_{b\tau}$    & 1.09 & 1.09  &1.09  & 1.09 & 1.08 \\
\hline
\hline
\end{tabular}
}
\caption{All points are consistent with the experimental constraints given in Section \ref{sec:scan} and the $b-\tau$ YU condition. All masses are given in GeV. Point 1 displays a solution for the gluino-neutralino coannihilation, while point 2 exemplifies a solution with stop-neutralino coannihilation. Point 3 depicts stau-neutralino coannihilation solution, and point 4 represents an A-funnel solution. Finally point 5 is for chargino-neutralino coannihilation.}
\label{table1}
\end{table}

Before concluding, we also present five benchmark points in Table \ref{table1} which summarize our findings. The points are all consistent with the experimental constraints given in Section \ref{sec:scan} and the $b-\tau$ YU condition. All masses are given in GeV. Point 1 displays a gluino-neutralino coannihilation solution with $m_{\tilde{g}}\sim m_{\tilde{\chi}_{1}^{0}} \simeq 1.1$ TeV. The Higgs boson mass is about 123 GeV with an uncertainty of 2 GeV or so. Point 2 exemplifies a solution for stop-neutralino coannihilation solution for $m_{\tilde{t}_{1}}\sim m_{\tilde{\chi}_{1}^{0}}\lesssim 0.9$ GeV. Point 3 depicts a stau-neutralino coannihilation solution in which the masses of stau and LSP neutralino lie within $10\%$ of each other, which is also the uncertainty in calculating the SUSY mass spectrum. This point also yields one of the highest cross-sections in DM scattering events. Point 4 represents an A-funnel solution. Finally point 5 is for chargino-neutralino coannihilation. Although all these four points are also solutions for chargino-neutralino coannihilation, in the fifth solution the desired relic abundance of the LSP neutralino is achieved exclusively through the chargino-neutralino coannihilation processes.

\section{Conclusion}
\label{sec:conc}

We have discussed $b-\tau$ YU in a class of SUSY $SU(5)$ GUTs in which SUSY breaking is transmitted through both gravity and anomaly mediation. This combination leads to non-universal gaugino masses at $M_{GUT}$, in contrast to unified gaugino masses in the simplest models. Due to the contributions from mirage mediation, $M_{3}$ is always lighter than $M_{1}$ and $M_{2}$, and consequently the gluino mass at the low scale is bounded from above at about 4 TeV (for the range of fundamental parameters considered). Considering the current mass bound on the gluino mass  $m_{\tilde{g}} \lesssim 2.1$ TeV ($m_{\tilde{g}} \lesssim 600$ GeV for $m_{\tilde{g}} \lesssim 1.1m_{\tilde{\chi}_{1}^{0}}$), the full mass range of the gluino will likely be tested at the HL-LHC and FCC experiments. $b-\tau$ YU solutions strongly depend on the threshold corrections to $y_{b}$, and these corrections satisfy $\delta_{y_{b}} \lesssim -0.3$.

We identified two different regions for the $\mu$ term, and its magnitude is less than about 3 TeV. This region yields a stop mass up to 5 TeV, and it is nearly degenerate in mass with the LSP neutralino in the 0.8 to 1.7 TeV range. Also, the stau mass can be realized up to about 5 TeV, and it can be approximately degenerate in mass with the LSP neutralino from about 2 to 3 TeV. In addition, we also found an A-funnel solution, which imply $m_{A}\approx 2m_{\tilde{\chi}_{1}^{0}}$, with $m_{\tilde{\chi}_{1}^{0}}\sim 700-900$ GeV. This region also favours coannihilation processes of LSP neutralino involving the stop and stau, and annihilation processes in which two LSP neutralinos annihilate into a CP-odd Higgs boson (A). The second region, on the other hand, arises for $m_{\tilde{g}}\lesssim 1.1m_{\tilde{\chi}_{1}^{0}}$. The $\mu-$term is rather large ($\gtrsim 20$ TeV), and the LSP neutralino is a bino-wino mixture. The gluino mass ($\sim 0.8-1.2$ TeV) is nearly degenerate with the LSP neutralino mass, and hence, the gluino-neutralino coannihilation processes take part in reducing the relic abundance of LSP neutralino down to ranges allowed by the current Planck measurements.

These two regions can be distinguished by considering the implications for the direct detection experiments. The first region with low $\mu$ yields Higgsino-like DM, whose scattering on the nucleus typically has a large cross-section. We find that such solutions are still allowed by the current LUX results, but they will be more severely tested by the LUX-Zeplin experiment. The second region, on the other hand, yield bino-wino DM, with a relatively smaller scattering cross-section.

 \section{Acknowledgement}
Q.S thanks the DOE for partial support provided under grant number DE-SC 0013880. CSU acknowledges  the National Academic Network and Information Center (ULAKBIM) of The Scientific and Technological Research Center of Turkey (TUBITAK), High Performance and Grid Computing Center (Truba Resources) for calculation of results presented in this paper.

\end{document}